\documentclass[english]{IEEEtran}
\usepackage[T1]{fontenc}
\usepackage[latin9]{inputenc}
\usepackage{color}
\usepackage{array}
\usepackage{rotating}
\usepackage{multirow}
\usepackage{amsmath}
\usepackage{amssymb}
\usepackage{stmaryrd}
\usepackage{graphicx}

\makeatletter

\providecommand{\tabularnewline}{\\}

\author{
    \IEEEauthorblockN{Abubakr O. Al-Abbasi, \textit{Student Member, IEEE}, Ridha Hamila, \textit{Senior Member, IEEE}, Waheed U. Bajwa, \textit{Senior Member, IEEE}, and Naofal Al-Dhahir, \textit{Fellow, IEEE}}

\thanks{This paper was made possible by grant number NPRP 06-070-2-024 from
the Qatar National Research Fund (a member of Qatar Foundation). The
statements made herein are solely the responsibility of the authors.

This work was presented in part at the 2015 IEEE Global Conference on Signal and Information Processing (GlobalSIP) [13].

Abubakr O. Al-Abbasi is with Purdue University, USA (e-mail: aalabbas@purdue.edu).

Ridha Hamila is with Qatar University, Qatar (e-mail: hamila@qu.edu.qa ).

Waheed U. Bajwa is with Rutgers University, The State University of New Jersey, USA (e-mail: waheed.bajwa@rutgers.edu).

Naofal Al-Dhahir is with the University of Texas at Dallas, USA
(e-mail: aldhahir@utdallas.edu).
}
}

\makeatother

\usepackage{babel}
\begin{document}

\title{Design and Analysis of Sparsifying Dictionaries for FIR MIMO Equalizers}
\maketitle
\begin{abstract}
In this paper, we propose a general framework that transforms the
problems of  designing sparse finite-impulse-response linear equalizers
and non-linear decision-feedback equalizers, for multiple antenna
systems, into the problem of sparsest-approximation of a vector in
different dictionaries. In addition, we investigate several choices
of the sparsifying dictionaries under this framework. Furthermore,
the worst-case coherences of these dictionaries, which determine their
sparsifying effectiveness, are analytically and/or numerically evaluated.
Moreover, we show how to reduce the computational complexity of the
designed sparse equalizer filters by exploiting the asymptotic equivalence
of Toeplitz and circulant matrices. Finally, the superiority of our
proposed framework over conventional methods is demonstrated through
numerical experiments.
\end{abstract}


\section{{\normalsize{}Introduction \label{sec:Introduction}}}

In single-carrier transmission over broadband channels, long finite
impulse response (FIR) equalizers are typically implemented at high
sampling rates to combat the channel's frequency selectivity. However,
implementation of such equalizers can be prohibitively expensive as
the design complexity of FIR equalizers grows proportional to the
square of the number of nonzero taps in the filter. Sparse equalization,
where only few nonzero coefficients are employed, is a widely-used
technique to reduce complexity at the cost of a tolerable performance
loss. Nevertheless, reliably determining the locations of these nonzero
coefficients is often very challenging. 

\textcolor{black}{Recently, sparse equalizers have been investigated
both from practical \cite{turboEq015,ionospheric015} and theoretical
\cite{sparseFilterDesign13,branchandBound13} perspectives to reduce
the implementation cost of long FIR filters. In \cite{turboEq015},
a direct-adaptive scheme is used for designing sparse FIR filters
for multi-channel turbo equalization in underwater acoustic communications.
However, the proposed approach is limited to the case of a single-input
linear equalizer. In \cite{ionospheric015}, the authors exploit sparsity
in ionospheric High Frequency (HF) communications systems by formulating
equalization at the HF receiver as a sparse signal recovery problem.
However, the resulting solution is not exactly sparse and an additional
heuristic optimization step is applied to further eliminate the small
nonzero entries. In \cite{sparseFilterDesign13}, a general optimization
problem for designing a sparse filter is formulated that involves
a quadratic constraint on filter performance. Nonetheless, the number
of iterations of the proposed algorithm becomes large as the desired
sparsity level of the filter increases. In addition, the approach
in \cite{sparseFilterDesign13} also involves inversion of a large
matrix in the case of a long channel impulse response (CIR). Sparse
filters can also be designed using integer programming methods \cite{branchandBound13}.
However, the design process can be computationally complex.}

In \cite{strongestTap}, the number of nonzero coefficients is reduced
by selecting only the significant taps of the equalizer. Nonetheless,
knowledge of the complete equalizer tap vector is still required,
which increases the computational complexity. In \cite{linearProgAlnOppen10},
an $\ell_{1}$-norm minimization problem is formulated to design a
sparse filter. However, since the resulting filter taps are not exactly
sparse, a thresholding step is required to force some of the nonzero
taps to $0$. An algorithm, called sparse chip equalizer, for finding
the locations of sparse equalizer taps is presented in \cite{sparseChipEqu},
but this approach assumes that the CIR itself is sparse. 

\textcolor{black}{In \cite{tapSelctDFE097}, an algorithm for designing
a decision feedback equalizer (DFE) is proposed, but the feedforward
filter (FFF) taps are designed to only equalize the channel taps having
the highest signal-to-noise ratios (SNRs). The number of the FFF and
feedback filter (FBF) taps are optimized in \cite{chosedfeTAPS}.
However, since no sparsity constraints are imposed on the design,
the final solution is not guaranteed to have low implementation complexity.
In \cite{mimoDFE}, multiple-input multiple-output (MIMO) equalizers
are optimally designed; however, the design complexity of the equalizers
is proportional to the product of the number of input and output streams.
The sparsity of some channel models, e.g., \cite{ITU-A} and \cite{TG3C},
is exploited in \cite{lee2004design} to further reduce the number
of equalizer taps. In \cite{vlachos2012stochastic}, a new matching-pursuit-type
algorithm for DFE adaptation is proposed and the direct-adaptive sparse
equalization problem is investigated from a compressive sensing perspective.
However, the algorithm in \cite{vlachos2012stochastic} exploits inherent
channel characteristics such as sparsity, i.e., the CIR is assumed
to have a large delay spread with only few dominant taps. In \cite{newDFW},
a framework for designing sparse FIR equalizers is proposed. Using
greedy algorithms, the proposed framework achieved better performance
than just choosing the largest taps of the minimum mean square error
(MMSE) equalizer, as in \cite{strongestTap}. However, this approach
involves inversion of large matrices and Cholesky factorization, whose
computational cost could be large for channels with large delay spreads.
In addition, no theoretical sparse approximation guarantees are provided. }

\textcolor{black}{In this paper, we develop a general framework for
the design of sparse FIR MIMO linear equalizers (LEs) and DFEs that
transforms the original problem into one of sparse approximation of
a vector using different dictionaries. The developed framework trivially
specializes to the case of single-input single-input (SISO) systems.
In both cases, the framework can then be used to find the sparsifying
dictionary that leads to the sparsest FIR filter subject to an approximation
constraint. Moreover, we investigate the coherence of the sparsifying
dictionaries that we propose as part of our analysis and identify
one dictionary that has small coherence. Then, we use simulations
to validate that the dictionary with the smallest coherence results
in the sparsest FIR design. For all design problems, we propose reduced-complexity
sparse FIR filter designs by exploiting the asymptotic equivalence
of Toeplitz and circulant matrices, where the matrix factorizations
involved in our design analysis can be carried out efficiently using
the fast Fourier transform (FFT) and inverse FFT with negligible performance
loss as the number of filter taps increases. Finally, numerical results
demonstrate the significance of our approach compared to conventional
sparse filter designs, e.g., in \cite{sigTaps} and \cite{strongestTap},
in terms of both performance and computational} complexity\footnote{\textcolor{black}{The design and analysis methods developed in this
paper are applicable to the wider class of FIR MMSE Wiener filters
(e.g., echo cancellers, noise rejection front-end filters, co-channel
interference canceller, etc.) and not limited only to equalizers.}}.

The remainder of this paper is organized as follows. After introducing
the system model in Section \ref{sec:Signal-Model}, we formulate
the sparse equalization problem for MIMO LEs and DFEs systems in Section
\ref{sec:Sparse-FIR-Equalization}. Our proposed unified framework
is described in Section \ref{sec:Proposed-sparse-approximation}.
Then, numerical results are presented in Section \ref{sec:Simulation-Results}.
Finally, the paper is concluded in Section \ref{sec:Conclusion-and-Future}.

\textbf{\textit{Notations}}: We use the following standard notation
in this paper: $\mbox{\ensuremath{\boldsymbol{I}}}_{N}$ denotes the
identity matrix of size $N$. Upper- and lower-case bold letters denote
matrices and vectors, respectively. Underlined upper-case bold letters,
e.g., $\boldsymbol{\underline{X}}$, denote frequency-domain vectors.
The notations $(.)^{-1},\,(.)^{*},\,(.)^{T}\mbox{ and }\,(.)^{H}$
denote the matrix inverse, the matrix (or element) complex conjugate,
the matrix transpose and the complex-conjugate transpose operations,
respectively. $\mbox{E\ensuremath{\left[.\right]}}$ denotes the expected
value operator. $\left\Vert .\right\Vert _{\ell}$ and $\left\Vert .\right\Vert _{F}$
denote the $\ell$-norm and Frobenius norm, respectively. $\otimes$
denotes the Kronecker product of matrices. The components of a vector
starting from $k_{1}$ and ending at $k_{2}$ are given as subscripts
to the vector separated by a colon, i.e., $\boldsymbol{x}_{k_{1}:k_{2}}.$ 

\begin{table}
\caption{\textcolor{black}{Channel equalization notation and key matrices used
in this paper.}\textit{\textcolor{black}{{} \label{tab:Channel-Equalization-notation}}}}

\textcolor{black}{\small{}}%
\begin{tabular}{|>{\raggedright}p{1.4cm}|>{\raggedright}p{3.2cm}|>{\raggedright}p{3cm}|}
\hline 
\textcolor{black}{\small{}Notation} & \textcolor{black}{\small{}Meaning} & \textcolor{black}{\small{}Size}\tabularnewline
\hline 
\textcolor{black}{\small{}$\boldsymbol{H}$} & \textcolor{black}{\footnotesize{}Channel matrix} & \textcolor{black}{\footnotesize{}$n_{o}N_{f}\times n_{i}\left(N_{f}+v\right)$}\tabularnewline
\hline 
\textcolor{black}{\small{}$\boldsymbol{R}_{xx}$} & \textcolor{black}{\footnotesize{}Input auto-correlation matrix } & \textcolor{black}{\scriptsize{}$n_{i}\left(N_{f}+v\right)\times\,\!\,n_{i}\left(N_{f}+v\right)$}\tabularnewline
\hline 
\textcolor{black}{\small{}$\boldsymbol{R}_{xy}$} & \textcolor{black}{\footnotesize{}Input-output cross-correlation matrix } & \textcolor{black}{\footnotesize{}$n_{i}\left(N_{f}+v\right)\times n_{o}\left(N_{f}\right)$}\tabularnewline
\hline 
\textcolor{black}{\small{}$\boldsymbol{R}_{yy}$} & \textcolor{black}{\footnotesize{}Output auto-correlation matrix } & \textcolor{black}{\footnotesize{}$n_{o}N_{f}\times n_{o}N_{f}$}\tabularnewline
\hline 
\textcolor{black}{\small{}$\boldsymbol{R}_{nn}$} & \textcolor{black}{\footnotesize{}Noise auto-correlation matrix } & \textcolor{black}{\footnotesize{}$n_{o}N_{f}\times n_{o}N_{f}$}\tabularnewline
\hline 
\textcolor{black}{\small{}$\boldsymbol{R}^{\perp}$} & \textcolor{black}{\footnotesize{}$\triangleq\boldsymbol{R}_{xx}-\boldsymbol{R}_{xy}\boldsymbol{R}_{yy}^{-1}\boldsymbol{R}_{yx}$ } & \textcolor{black}{\footnotesize{}$n_{i}\left(N_{f}+v\right)\times\,\,\!n_{i}\left(N_{f}+v\right)$}\tabularnewline
\hline 
\textcolor{black}{\small{}$\boldsymbol{W}$} & \textcolor{black}{\footnotesize{}FFF matrix cofficients } & \textcolor{black}{\footnotesize{}$n_{o}N_{f}\times n_{i}$}\tabularnewline
\hline 
\textcolor{black}{\small{}$\boldsymbol{B}$} & \textcolor{black}{\footnotesize{}FBF matrix cofficients } & \textcolor{black}{\footnotesize{}$n_{i}\left(N_{f}+v\right)\times n_{i}$}\tabularnewline
\hline 
\end{tabular}{\small \par}

\end{table}

\section{{\normalsize{}System Model \label{sec:Signal-Model}}}

We consider a linear time-invariant MIMO inter-symbol interference
(ISI) channel with $n_{i}$ inputs and $n_{o}$ outputs\textit{\textcolor{blue}{{}
}}\textcolor{black}{(the key matrices used in this paper are summarized
in Table \ref{tab:Channel-Equalization-notation}). The received sample
at the $r^{th}$ output antenna $\left(1\leq r\leq n_{o}\right)$
at time $k$ can be expressed as }

\textcolor{black}{
\begin{equation}
y_{k}^{\left(r\right)}=\sum_{i=1}^{n_{i}}\sum_{l=0}^{v^{\left(i,r\right)}}\boldsymbol{h}_{l}^{\left(i,r\right)}\boldsymbol{x}_{k-l}^{\left(i\right)}+n_{k}^{\left(r\right)}\,,
\end{equation}
where $y_{k}^{\left(r\right)}$ is the $r^{th}$ channel output, $\boldsymbol{h}_{l}^{\left(i,r\right)}$
is the CIR between the $i^{th}$ input and the $r^{th}$ output whose
memory length is $v^{\left(i,r\right)}$, and $n_{k}^{\left(r\right)}$
is the noise at the $r^{th}$ output antenna.  The received samples
from all $n_{o}$ channel outputs at sample time $k$ are grouped
into a $n_{o}\times1$ column vector $\boldsymbol{y}_{k}$ as follows: }

\textcolor{black}{
\begin{equation}
\boldsymbol{y}_{k}=\sum_{l=0}^{v}\boldsymbol{H}_{l}\boldsymbol{x}_{k-l}+\boldsymbol{n}_{k}\,,\label{eq:y_k}
\end{equation}
}where $\boldsymbol{H}_{l}$ is the $l^{th}$ channel matrix coefficient
of dimension $\left(n_{o}\times n_{i}\right)$, and $\boldsymbol{x}_{k-l}$
is size $n_{i}\times1$ input vector at time $k-l$. The parameter
$v$ is the maximum order of all of the $n_{o}n_{i}$ CIRs, i.e.,
$v = \max_{(i,r)} v^{(i,r)}$. Over a block of $N_{f}$ output samples,
the input-output relation in (\ref{eq:y_k}) can be written compactly
as 

\begin{equation}
\boldsymbol{y}_{k:k-N_{f}+1}=\boldsymbol{H}\,\boldsymbol{x}_{k:k-N_{f}-v+1}+\boldsymbol{n}_{k:k-N_{f}+1}\,,\label{eq:y_Hx_n}
\end{equation}
where $\boldsymbol{y}_{k:k-N_{f}+1},\,\boldsymbol{x}_{k:k-N_{f}-v+1}$
and $\boldsymbol{n}_{k:k-N_{f}+1}$ are column vectors grouping the
received, transmitted and noise samples, respectively. \textcolor{black}{Recall
that $\boldsymbol{y}_{k:k-N_{f}+1}$ is a vector of length $n_{o}N_{f}$,
i.e., $\boldsymbol{y}=\left[\begin{array}{cccc}
\boldsymbol{y}_{k} & \boldsymbol{y}_{k-1} & \ldots & \boldsymbol{y}_{k-N_{f}+1}\end{array}\right]^{T}$}. Additionally, $\boldsymbol{H}$ is a block Toeplitz matrix whose
first block row is formed by $\{\boldsymbol{H}_{l}\}_{l=0}^{l=v}$
followed by zero matrices. It is useful, as will be shown in the sequel,
to define the output auto-correlation and the input-output cross-correlation
matrices based on the block of length $N_{f}$. Using (\ref{eq:y_Hx_n}),
the $n_{i}(N_{f}+v)\times n_{i}(N_{f}+v)$ input correlation and the
$n_{o}N_{f}\times n_{o}N_{f}$ noise correlation matrices are, respectively,
defined by $\boldsymbol{R}_{xx}\triangleq E\left[\boldsymbol{x}_{k:k-N_{f}-v+1}\boldsymbol{x}_{k:k-N_{f}-v+1}^{H}\right]\mbox{ and }\boldsymbol{R}_{nn}\triangleq E\left[\boldsymbol{n}_{k:k-N_{f}+1}\boldsymbol{n}_{k:k-N_{f}+1}^{H}\right]$.
Both the input and noise processes are assumed to be white; hence,
their auto-correlation matrices are assumed to be (multiples of) the
identity matrix, i.e., $\boldsymbol{R}_{xx}=\boldsymbol{I}_{n_{i}(N_{f}+v)}$
and $\boldsymbol{R}_{nn}=\frac{1}{SNR}\boldsymbol{I}_{n_{o}N_{f}}$.
Moreover, the output-input cross-correlation and the output auto-correlation
matrices are, respectively, defined as

\begin{eqnarray}
\!\!\boldsymbol{R}_{yx} & \!\!\triangleq & \!\!E\left[\!\boldsymbol{y}_{k:k-N_{f}+1}\boldsymbol{x}_{k:k-N_{f}-v+1}^{H}\!\right]\!\!=\boldsymbol{H}\boldsymbol{R}_{xx}\,,\,\mbox{and}\\
\!\!\boldsymbol{R}_{yy} & \!\!\triangleq & \!\!E\left[\!\boldsymbol{y}_{k:k-N_{f}+1}\boldsymbol{y}_{k:k-N_{f}+1}^{H}\!\right]\!\!=\!\boldsymbol{H}\boldsymbol{R}_{xx}\boldsymbol{H}^{H}\!\!+\!\boldsymbol{R}_{nn}.\label{eq:R_yy_def}
\end{eqnarray}

\section{Sparse FIR Equalization \label{sec:Sparse-FIR-Equalization}}

In this section, we formulate the sparse FIR equalizer design problems
for MIMO LEs and DFEs.

\subsection{Sparse FIR MIMO LE\label{subsec:Sparse-FIR-MIMO-LE}}

\textcolor{black}{The received samples are passed through a MIMO FIR
filter of length $n_{o}N_{f}$ for equalization. Define the $k^{th}$
equalization error sample vector in the MIMO setting as \cite{mimoDFE}}

\textcolor{black}{
\begin{equation}
\boldsymbol{e}_{k}=\left[\begin{array}{ccccc}
e_{k,1} & e_{k,2} & \ldots & \ldots & e_{k,n_{i}}\end{array}\right]^{T}\,,
\end{equation}
where $e_{k,i}$ is the equalization error of the $i^{th}$ input
stream. The resulting $k^{th}$ error sample for the $i^{th}$ input
stream can be expressed as \cite{mimoDFE}}

\textcolor{black}{
\begin{equation}
e_{k,i}=x_{k-\Delta,i}-\hat{x}_{k}=x_{k-\Delta,i}-\boldsymbol{w}_{i}^{H}\boldsymbol{y}_{k:k-N_{f}+1}\,,
\end{equation}
where $\Delta$ is the decision delay, typically }\textcolor{black}{\small{}$0\leq\Delta\leq N_{f}+v-1$}\textcolor{black}{,
and $\boldsymbol{w}_{i}$ denotes the equalizer taps vector for the
$i^{th}$ input stream whose dimension is $n_{o}N_{f}\times1$. The
MSE of $e_{k,i}$, i.e., $\ensuremath{\xi}_{i}\left(\boldsymbol{w}_{i}\right)$,
for the $i^{th}$ input stream can be written as}

\textcolor{black}{
\begin{equation}
\ensuremath{\xi}_{i}\left(\boldsymbol{w}_{i}\right)=\xi_{m,i}+\underbrace{(\boldsymbol{w}_{i}-\boldsymbol{R}_{yy}^{-1}\boldsymbol{r}_{\Delta,i})^{H}\boldsymbol{R}_{yy}(\boldsymbol{w}_{i}-\boldsymbol{R}_{yy}^{-1}\boldsymbol{r}_{\Delta,i})}_{\triangleq\xi_{ex,i}(\boldsymbol{w}_{i})}\,,\label{eq:error_mimo-LEs}
\end{equation}
where $\xi_{m,i}\triangleq\ensuremath{\varepsilon_{x,i}}-\boldsymbol{r}_{\Delta,i}^{H}\boldsymbol{R}_{yy}^{-1}\boldsymbol{r}_{\Delta,i}$,
$\varepsilon_{x,i}\triangleq E\left[x_{k-\Delta,i}^{2}\right]$, $\boldsymbol{r}_{\Delta,i}^{H}=\boldsymbol{R}_{yx}\boldsymbol{1}_{\Delta,i}$,
and $\boldsymbol{1}_{\Delta,i}$ is the $(n_{i}\Delta+i)$-th column
of $\boldsymbol{I}_{n_{i}(N_{f}+v)}$. Clearly, the optimum choice
for $\boldsymbol{w}_{i}$, in the MMSE sense, is the complex non-sparse
solution: $\boldsymbol{w}_{opt,i}=\boldsymbol{R}_{yy}^{-1}\boldsymbol{r}_{\Delta,i}.$
However, in general, $\boldsymbol{w}_{opt.i}$ is not sparse and its
implementation complexity increases proportional to $(n_{o}N_{f})^{2}$,
which can be computationally expensive \cite{DCProakis}. However,
any choice for $\boldsymbol{w}_{i}$ other than $\boldsymbol{w}_{opt,i}$
increases $\xi_{i}(\boldsymbol{w}_{i})$, which results in performance
loss. This suggests that we can use the excess error $\xi_{ex,i}(\boldsymbol{w}_{i})$
as a design constraint to achieve a desirable performance-complexity
tradeoff. Specifically, we formulate the following problem for the
design of a sparse FIR MIMO LE}

\textcolor{black}{
\begin{eqnarray}
\widehat{\boldsymbol{w}}_{s,i} & \triangleq & \underset{\boldsymbol{w}_{i}\in\mathbb{C}^{n_{o}N_{f}}}{\mbox{arg}\mbox{min}}\,\,\left\Vert \boldsymbol{w}_{i}\right\Vert _{0}\,\,\,\,\mbox{subject to}\,\,\,\,\,\xi_{ex,i}(\boldsymbol{w}_{i})\leq\delta_{eq,i}\,,\nonumber \\
\label{eq:opt_prob1}
\end{eqnarray}
where $\left\Vert \boldsymbol{w}_{i}\right\Vert _{0}$ is the number
of nonzero elements in its argument and $\delta_{eq,i}$ can be chosen
as a function of the noise variance. To solve (\ref{eq:opt_prob1}),
we propose a general framework presented in the sequel to sparsely
design FIR MIMO LEs such that the performance loss does not exceed
a pre-specified limit. We conclude this section by pointing out that
the setup in (\ref{eq:opt_prob1}) can be easily specialized to the
case of sparse FIR SISO LEs.}

\subsection{Sparse FIR MIMO DFE\label{subsec:Sparse-FIR-MIMO-DFE}}

\textcolor{black}{The FIR MIMO-DFE consists of two filters: a FFF
matrix \cite{mimoDFE}}

\textcolor{black}{\small{}
\begin{equation}
\boldsymbol{W}^{H}\triangleq\left[\begin{array}{cccc}
\boldsymbol{W}_{0}^{H} & \boldsymbol{W}_{1}^{H} & \ldots & \boldsymbol{W}_{N_{f}-1}^{H}\end{array}\right]\,,
\end{equation}
}\textcolor{black}{with $N_{f}$ matrix taps $\boldsymbol{W}_{i}^{H}$,
each of size $n_{o}\times n_{i}$, and a FBF matrix equal to }

\textcolor{black}{\small{}
\begin{equation}
\widetilde{\boldsymbol{B}}^{H}=\left[\begin{array}{cccc}
\widetilde{\boldsymbol{B}}_{0}^{H} & \widetilde{\boldsymbol{B}}_{1}^{H} & \ldots & \widetilde{\boldsymbol{B}}_{N_{b}}^{H}\end{array}\right]\,,
\end{equation}
}\textcolor{black}{where each $\widetilde{\boldsymbol{B}}_{i}^{H}$
has $\left(N_{b}+1\right)$ taps with size of $n_{i}\times n_{i}$.
Therefore, $\boldsymbol{W}_{i}$ and $\widetilde{\boldsymbol{B}}_{i}$
have the forms }

\textit{\textcolor{black}{
\begin{equation}
\boldsymbol{W}_{i}=\left[\begin{array}{ccc}
w_{i}^{\left(1,1\right)} & \ldots & w_{i}^{\left(1,n_{i}\right)}\\
\vdots & \ldots & \vdots\\
w_{i}^{\left(n_{o},1\right)} &  & w_{i}^{\left(n_{o},n_{i}\right)}
\end{array}\right]
\end{equation}
}}

\textit{\textcolor{black}{
\begin{equation}
\widetilde{\boldsymbol{B}}_{i}=\left[\begin{array}{ccc}
b_{i}^{\left(1,1\right)} & \ldots & b_{i}^{\left(1,n_{i}\right)}\\
\vdots & \ldots & \vdots\\
b_{i}^{\left(n_{o},1\right)} &  & b_{i}^{\left(n_{o},n_{i}\right)}
\end{array}\right]
\end{equation}
}}

\textcolor{black}{By defining the size $n_{i}\times n_{i}(N_{f}+v)$
matrix}\textcolor{black}{\small{} $\boldsymbol{B}^{H}=\left[\begin{array}{cc}
\boldsymbol{0}_{n_{i}\times n_{i}\Delta} & \widetilde{\boldsymbol{B}}^{H}\end{array}\right]$}\textcolor{black}{, where $0\leq\Delta\leq N_{f}+v-1$ is the decision
delay that satisfies the condition $\left(\Delta+N_{b}+1\right)=\left(N_{f}+v\right)$,
it was shown in \cite{mimoDFE} that the MSE of the error vector at
time $k$, i.e., $\boldsymbol{E}_{k}=\boldsymbol{B}^{H}\boldsymbol{x}_{k:k-N_{f}-v+1}-\boldsymbol{W}^{H}\boldsymbol{y}_{k:k-N_{f}+1}$,
is given by \cite{mimoCS,csMIMOSparse}}

\textcolor{black}{\small{}
\begin{eqnarray}
\!\!\!\xi\left(\boldsymbol{B},\boldsymbol{W}\right) & \!\!\!=\!\!\! & \ensuremath{\underbrace{\mbox{ Trace}\left\{ \boldsymbol{B}^{H}\boldsymbol{R}^{\perp}\boldsymbol{B}\vphantom{\boldsymbol{S}^{H}\boldsymbol{R}_{yy}}\right\} }_{\triangleq\,\xi_{m}\left(\boldsymbol{B}\right)}+\underbrace{\mbox{\ensuremath{\mbox{Trace}\left\{ \boldsymbol{S}^{H}\boldsymbol{R}_{yy}\boldsymbol{S}\right\} }}}_{\ensuremath{\triangleq\,\xi_{ex}(\boldsymbol{W},\boldsymbol{B})}}\,,}\label{eq:MSE-MIMO-BRB}
\end{eqnarray}
}\textcolor{black}{where $\boldsymbol{R}^{\perp}\triangleq\boldsymbol{R}_{xx}-\boldsymbol{R}_{xy}\boldsymbol{R}_{yy}^{-1}\boldsymbol{R}_{yx}$
and}\textcolor{black}{\small{} }\textcolor{black}{$\boldsymbol{S}^{H}\triangleq\!\boldsymbol{W}^{H}\!\!-\!\!\boldsymbol{B}^{H}\boldsymbol{R}_{xy}\boldsymbol{R}_{yy}^{-1}.$
The second term of the MSE in (\ref{eq:MSE-MIMO-BRB}) is equal to
zero for the case of the optimum FFF matrix filter coefficients, i.e.,}\textcolor{black}{\small{}
}\textcolor{black}{$\boldsymbol{W}^{H}\!\!=\!\!\boldsymbol{B}^{H}\boldsymbol{R}_{xy}\boldsymbol{R}_{yy}^{-1}$,
and the resulting MSE can then be expressed as follows}\footnote{\textcolor{black}{We express $\boldsymbol{R}^{\perp}$ as $\boldsymbol{A}_{\perp}^{H}\boldsymbol{A}_{\perp}$,
where $\boldsymbol{A}_{\perp}$ is the square-root matrix of $\boldsymbol{R}^{\perp}$
in the spectral-norm sense and results from Cholesky or eigen decompositions
\cite{matAnalysis}.}}\textcolor{black}{\small{} }\textcolor{black}{$\left(\mbox{defining}\boldsymbol{\,R}^{\perp}\triangleq\boldsymbol{A}_{\perp}^{H}\boldsymbol{A}_{\perp}\right)$}

\textcolor{black}{\small{}
\begin{eqnarray}
\xi_{m}\left(\boldsymbol{B}\right) & \!\!\!\!=\!\!\!\! & \mbox{Trace}\left\{ \boldsymbol{B}^{H}\boldsymbol{A}_{\perp}^{H}\boldsymbol{A}_{\perp}\boldsymbol{B}\right\} =\left\Vert \boldsymbol{A}_{\perp}\boldsymbol{B}\vphantom{A^{H}}\right\Vert _{F}^{2}\nonumber \\
 & \!\!\!\!=\!\!\!\! & \left\Vert \boldsymbol{A}_{\perp}\begin{array}{cccc}
\boldsymbol{b}^{\left(1\right)} & \boldsymbol{A}_{\perp}\boldsymbol{b}^{\left(2\right)} & \ldots\ldots & \boldsymbol{A}_{\perp}\boldsymbol{b}^{\left(n_{i}\right)}\end{array}\right\Vert _{F}^{2}\nonumber \\
 & = & \!\!\!\left\Vert \boldsymbol{A}_{\perp}\!\!\begin{array}{c}
\boldsymbol{b}^{\left(1\right)}\!\!\end{array}\right\Vert _{2}^{2}+\!\!\left\Vert \boldsymbol{A}_{\perp}\!\!\begin{array}{c}
\boldsymbol{b}^{\left(2\right)}\!\!\end{array}\right\Vert _{2}^{2}\!+\cdots\cdots+\!\!\left\Vert \boldsymbol{A}_{\perp}\!\!\begin{array}{c}
\boldsymbol{b}^{\left(n_{i}\right)}\!\!\end{array}\right\Vert _{2}^{2}\nonumber \\
\label{eq:FBF_design_B}
\end{eqnarray}
}\textcolor{black}{where $\boldsymbol{b}^{(i)}$ is the $i^{th}$
column of $\boldsymbol{B}$. Hence, to compute the FBF matrix filter
taps $\boldsymbol{B}$ that minimize $\xi_{m}\left(\boldsymbol{B}\right)$,
we minimize $\xi_{m}\left(\boldsymbol{B}\right)$ under the identity
tap constraint (ITC), i.e., we restrict $\boldsymbol{B}_{0}$ to
be equal to the identity matrix, i.e., $\boldsymbol{B}_{0}=\boldsymbol{I}_{n_{i}}$.
Towards this goal, we rewrite $\xi_{m}\left(\boldsymbol{B}\right)$
as follows}

\textcolor{black}{
\begin{equation}
\xi_{m}\left(\boldsymbol{B}\right)=\sum_{i=1}^{n_{i}}\left\Vert \boldsymbol{A}_{\perp}^{(:\setminus n_{i}\Delta+i)}\begin{array}{c}
\boldsymbol{b}^{\left(i\setminus n_{i}\Delta+i\right)}+\boldsymbol{a}_{n_{i}\Delta+i}\end{array}\right\Vert _{2}^{2}\,,
\end{equation}
where $\boldsymbol{A}_{\perp}^{(:\setminus n_{i}\Delta+i)}$ is formed
by all columns of $\boldsymbol{A}_{\perp}$ except the $\left(n_{i}\Delta+i\right)^{th}$
column, i.e., $\boldsymbol{a}_{n_{i}\Delta+i}$, and $\boldsymbol{b}^{\left(i\setminus n_{i}\Delta+i\right)}$
is formed by all elements of $\begin{array}{c}
\boldsymbol{b}^{\left(i\right)}\end{array}$ except the $\left(n_{i}\Delta+i\right)^{th}$ entry that is forced
to have unit value. Then, we formulate the following problem for the
design of sparse FBF matrix filter taps $\boldsymbol{B}$}

\textcolor{black}{
\begin{equation}
\begin{array}{c}
\,\,\,\widehat{\boldsymbol{b}}^{\left(i\setminus n_{i}\Delta+i\right)}\triangleq\underset{}{\mbox{\mbox{arg}\mbox{min}}}\,\left\Vert \boldsymbol{b}^{\left(i\setminus n_{i}\Delta+i\right)}\right\Vert _{0}\,\,\,\,\,\,\mbox{subject to}\\
\left\Vert \boldsymbol{A}_{\perp}^{(:\setminus n_{i}\Delta+n_{i})}\begin{array}{c}
\boldsymbol{b}^{\left(i\setminus n_{i}\Delta+i\right)}+\boldsymbol{a}_{n_{i}\Delta+i}\mbox{\!\!\!}\end{array}\right\Vert _{2}^{2}\leq\gamma_{eq,i}\,,
\end{array}\label{eq:mimoDFE-FBF}
\end{equation}
where $\widehat{\boldsymbol{b}}^{\left(.\right)}$ is the estimate
of $\boldsymbol{b}^{\left(.\right)}$. Once $\widehat{\boldsymbol{b}}^{\left(i\setminus n_{i}\Delta+i\right)}\,,\forall i\in n_{i}$,
is calculated, we insert the identity matrix $\boldsymbol{B}_{0}$
in the first location of $\boldsymbol{B}$ to form the sparse FBF
matrix coefficients, i.e., $\boldsymbol{B}_{s}$. Note that $\gamma_{eq,i}$
can be used to provide different quality of service (QoS) levels,
with small values assigned to users/streams that demand high QoS levels.
Then, the optimum FFF matrix taps (in the MMSE sense) are determined
from (\ref{eq:MSE-MIMO-BRB}) to be }

\textcolor{black}{
\begin{equation}
\boldsymbol{W}_{opt}=\boldsymbol{R}_{yy}^{-1}\boldsymbol{R}_{yx}\boldsymbol{B}_{s}=\boldsymbol{R}_{yy}^{-1}\boldsymbol{\overline{\beta}}\,.\label{eq:wopt_bopt_mimo_sparse}
\end{equation}
}

\textcolor{black}{Since $\boldsymbol{W}_{opt}$ is not sparse in
general, we further propose a sparse implementation for the FFF matrix
as follows. After computing $\boldsymbol{B}_{s}$, the MSE will be
a function only of $\boldsymbol{W}$ and can be expressed as $\left(\mbox{defining}\boldsymbol{\,R}_{yy}\triangleq\boldsymbol{A}_{y}^{H}\boldsymbol{A}_{y}\right)$}

\textcolor{black}{\small{}
\begin{eqnarray}
\xi\left(\boldsymbol{B}_{s},\boldsymbol{W}\right) & = & \xi_{m}\left(\boldsymbol{B}_{s}\right)+\nonumber \\
 &  & \ensuremath{\mbox{\ensuremath{\mbox{Trace}\left\{ \!\left(\!\boldsymbol{W}^{H}\!\!-\!\!\boldsymbol{\overline{\beta}}^{H}\boldsymbol{R}_{yy}^{-1}\right)\boldsymbol{A}_{y}^{H}\boldsymbol{A}_{y}\left(\!\boldsymbol{W}\!\!-\!\!\boldsymbol{R}_{yy}^{-1}\boldsymbol{\overline{\beta}}\right)\!\right\} }}}\nonumber \\
 & = & \xi_{m}\left(\boldsymbol{B}_{s}\right)+\underbrace{\ensuremath{\left\Vert \boldsymbol{A}_{y}\boldsymbol{W}-\boldsymbol{A}_{y}^{-H}\boldsymbol{\overline{\beta}}\right\Vert _{F}^{2}}}_{\,\ensuremath{\triangleq\,\xi_{ex}(\boldsymbol{W})}}.\label{eq:MSE-MIMO-DFE-FFF}
\end{eqnarray}
}{\small \par}

\textcolor{black}{By minimizing $\ensuremath{\xi_{ex}(\boldsymbol{W})}$,
we further minimize the MSE. This is achieved by a reformulation for
$\ensuremath{\xi_{ex}(\boldsymbol{W})}$ to get a vector form of $\boldsymbol{W}$,
 as follows}

\textcolor{black}{\small{}
\begin{equation}
\ensuremath{\xi_{ex}(\overline{\boldsymbol{w}}_{f})=}\left\Vert \underbrace{\left(\boldsymbol{I}_{n_{i}}\otimes\boldsymbol{A}_{y}^{H}\right)}_{\overline{\boldsymbol{\varPsi}}}\mbox{\ensuremath{\underbrace{\textbf{vec}\left(\boldsymbol{W}\right)\vphantom{\left(\boldsymbol{I}_{n_{i}}\otimes\boldsymbol{A}_{y}^{H}\right)}}_{\overline{\boldsymbol{w}}_{f}}-}\ensuremath{\underbrace{\textbf{vec}(\boldsymbol{A}_{y}^{-H}\boldsymbol{\overline{\beta}})\vphantom{\left(\boldsymbol{I}_{n_{i}}\otimes\boldsymbol{A}_{y}^{H}\right)}}_{\overline{\boldsymbol{a}}_{y}}}}\right\Vert _{2}^{2}\,,
\end{equation}
}\textcolor{black}{where $\mbox{\textbf{vec}}$ is an operator that
maps a $n\times n$ matrix to a vector by stacking the columns of
the matrix. Afterward, we solve the following problem to compute the
FFF matrix filter taps}

\textit{\textcolor{black}{
\begin{equation}
\widehat{\overline{\boldsymbol{w}}}_{s,f}\triangleq\underset{}{\mbox{\mbox{arg}\mbox{min}}}\,\left\Vert \overline{\boldsymbol{w}}_{f}\right\Vert _{0}\,\,\,\mbox{subject to}\,\,\,\,\,\,\ensuremath{\xi_{ex}(\overline{\boldsymbol{w}}_{f})}\leq\overline{\gamma}_{eq}\,,\label{eq:MIMO-DFE-FFF}
\end{equation}
}}\textcolor{black}{where $\overline{\gamma}_{eq}>0$ is used to control
the performance-complexity tradeoff. We conclude this section by noting
that the FIR LEs follow as a special case of the FIR DFEs by setting
$\boldsymbol{B}_{0}=\boldsymbol{I}_{n_{i}}$ and $\boldsymbol{B}_{\ell}=\boldsymbol{0}_{n_{i}\times n_{i}}$,
$0\leq\ell\leq N_{b}$. In addition, the MSE matrix of DFE is a weighted
version of that of the LE \cite{7093182}. Moreover, the setup in
(\ref{eq:mimoDFE-FBF}) and (\ref{eq:MIMO-DFE-FFF}) can be easily
specialized to the case of sparse FIR SISO DFEs by setting the numbers
of inputs and outputs to one.}

\section{\textcolor{black}{Proposed Sparse Approximation Framework\label{sec:Proposed-sparse-approximation}}}

\textcolor{black}{Unlike earlier works, including the one by one of
the co-authors \cite{newDFW}, we provide a general framework for
designing sparse FIR filters, for multiple antenna systems, that can
be considered as the problem of sparse approximation using different
dictionaries. Mathematically, this framework poses the FIR filter
design problem as follows}

\textcolor{black}{
\begin{equation}
\widehat{\boldsymbol{z}}_{s}\triangleq\underset{\boldsymbol{z}}{\mbox{\mbox{arg}\mbox{min}}}\,\left\Vert \boldsymbol{z}\right\Vert _{0}\,\,\,\mbox{subject to}\,\,\,\left\Vert \boldsymbol{K}\left(\boldsymbol{\varPhi}\boldsymbol{z}-\boldsymbol{d}\right)\right\Vert _{2}^{2}\leq\epsilon\,,\label{eq:propFW}
\end{equation}
where $\boldsymbol{\varPhi}$ is the dictionary that will be used
to sparsely approximate $\boldsymbol{d}$, while $\boldsymbol{K}$
is a known matrix and $\boldsymbol{d}$ is a known data vector, both
of which change depending upon the sparsifying dictionary $\boldsymbol{\varPhi}$.
Notice that $\widehat{\boldsymbol{z}}_{s}$ corresponds to one of
the elements in $\{\widehat{\boldsymbol{w}}_{s,i},\,\widehat{\boldsymbol{b}}^{\left(.\right)},\,\widehat{\boldsymbol{\overline{w}}}_{s,f}\}$
and $\epsilon$ is the corresponding element in $\left\{ \delta_{eq,i},\,\gamma_{eq,i},\,\overline{\gamma}_{eq}\right\} $.
For all design problems, we perform the suitable transformation 
to reduce the problem to the one shown in (\ref{eq:propFW}). For
instance, we complete the square in (\ref{eq:opt_prob1}) to reduce
it to the formulation given in (\ref{eq:propFW}). Hence, one can
use any factorization for $\boldsymbol{R}_{yy}$, e.g., in (\ref{eq:error_mimo-LEs})
or (\ref{eq:MSE-MIMO-BRB}), and $\boldsymbol{R}^{\perp}$, e.g.,
in (\ref{eq:MSE-MIMO-BRB}), to formulate a sparse approximation problem.
Using the Cholesky or eigen decomposition for $\boldsymbol{R}_{yy},$
and $\boldsymbol{R}^{\perp}$, we will have different choices for
$\boldsymbol{K}$, $\boldsymbol{\varPhi}$ and $\boldsymbol{d}$.
For instance, by defining the Cholesky factorization \cite{matAnalysis}
of $\boldsymbol{R}^{\perp}$, in (\ref{eq:FBF_design_B}), as $\boldsymbol{R}^{\perp}\triangleq\boldsymbol{L}_{\perp}\boldsymbol{L}_{\perp}^{H}$,
or in the equivalent form $\boldsymbol{R}^{\perp}\triangleq\boldsymbol{P}_{\perp}\boldsymbol{\Sigma}_{\perp}\boldsymbol{P}_{\perp}^{H}=\boldsymbol{\varOmega}_{\perp}\boldsymbol{\varOmega}_{\perp}^{H}$
(where $\boldsymbol{L}_{\perp}$ is a lower-triangular matrix, $\boldsymbol{P}_{\perp}$
is a lower-unit-triangular (unitriangular) matrix and $\boldsymbol{\Sigma}_{\perp}$
is a diagonal matrix) and assuming $n_{i}=1$ (in which matrix $\boldsymbol{B}$
reduces to a vector $\boldsymbol{b}$), the problem in (\ref{eq:propFW})
can, respectively, take one of the forms shown below \cite{ourFW_CSE_TIR}}

\textcolor{black}{\small{}
\begin{eqnarray}
 & \underset{\boldsymbol{b}\in\mathbb{C}^{N_{f}+v-1}}{\mbox{min}}\,\,\left\Vert \boldsymbol{b}\right\Vert _{0}\mbox{\,\,\,\,\mbox{s.t. }\,\,\,\,\ensuremath{\left\Vert \left(\widetilde{\boldsymbol{L}}_{\perp}^{H}\,\widetilde{\boldsymbol{b}}+\boldsymbol{l}_{\Delta+1}\right)\right\Vert _{2}^{2}\leq\gamma_{eq,1}\,},}\label{eq:ex_R_delta_1}\\
 & \underset{\boldsymbol{b}\in\mathbb{C}^{N_{f}+v-1}}{\mbox{min}}\,\,\left\Vert \boldsymbol{b}\right\Vert _{0}\mbox{\,\,\,\,\mbox{s.t. }\,\,\,\,\ensuremath{\left\Vert \left(\widetilde{\boldsymbol{\varOmega}}_{\perp}^{H}\,\widetilde{\boldsymbol{b}}+\boldsymbol{p}_{\Delta+1}\right)\right\Vert _{2}^{2}\leq\gamma_{eq,1}\,}.}\label{eq:ex_R_delta_2}
\end{eqnarray}
}\textcolor{black}{{} Note that $\widetilde{\boldsymbol{L}}_{\perp}^{H}$$\left(\widetilde{\boldsymbol{\varOmega}}_{\perp}^{H}\right)$
is formed by all columns of $\boldsymbol{L}_{\perp}^{H}$$\left(\boldsymbol{\varOmega}_{\perp}^{H}\right)$
except the $\left(\Delta+1\right)^{th}$ column, $\boldsymbol{l}_{\Delta+1}$$\left(\boldsymbol{p}_{\Delta+1}\right)$
is the $\left(\Delta+1\right)^{th}$ column of $\boldsymbol{L}_{\perp}^{H}$$\left(\boldsymbol{\varOmega}_{\perp}^{H}\right)$,
and $\widetilde{\boldsymbol{b}}$ is formed by all entries of $\boldsymbol{b}$
except the $\left(\Delta+1\right)^{th}$ unity entry. Similarly, by
writing the Cholesky factorization of $\boldsymbol{R}_{yy}$ in (\ref{eq:error_mimo-LEs})
as $\boldsymbol{R}_{yy}\triangleq\boldsymbol{L}_{y}\boldsymbol{L}_{y}^{H}$
or the eigen decomposition of $\boldsymbol{R}_{yy}$ as $\boldsymbol{R}_{yy}\triangleq\boldsymbol{U}_{y}\boldsymbol{D}_{y}\boldsymbol{U}_{y}^{H}$,
 we can formulate the problem in (\ref{eq:propFW}) as follows}

\textcolor{black}{\small{}
\begin{eqnarray}
\!\!\!\!\!\!\!\!\!\! & \!\!\!\!\underset{\boldsymbol{w}_{i}\in\mathbb{C}^{n_{o}N_{f}}}{\mbox{min}}\left\Vert \boldsymbol{w}_{i}\right\Vert _{0}\mbox{\,\,\,\,\mbox{s.t. }\,\,\,\,\ensuremath{\left\Vert (\boldsymbol{L}_{y}^{H}\boldsymbol{w}_{i}-\boldsymbol{L}_{y}^{-1}\boldsymbol{r}_{\Delta,i})\right\Vert _{2}^{2}\leq\delta_{eq,i}\,},}\label{eq:L_y_h}\\
\!\!\!\!\!\!\!\! & \!\!\!\!\!\underset{\boldsymbol{w}_{i}\in\mathbb{C}^{n_{o}N_{f}}}{\mbox{min}}\!\!\left\Vert \boldsymbol{w}_{i}\right\Vert _{0}\mbox{\,\mbox{s.t. }\,\ensuremath{\left\Vert \boldsymbol{D}_{y}^{\frac{1}{2}}\boldsymbol{U}_{y}^{H}\boldsymbol{w}_{i}\!-\!\boldsymbol{D}_{y}^{-\frac{1}{2}}\boldsymbol{U}_{y}^{H}\boldsymbol{r}_{\Delta,i}\right\Vert _{2}^{2}\!\!\!\leq\!\delta_{eq,i},\,\mbox{and}\!\!}}\label{eq:U_telda_y}\\
\!\!\!\!\!\!\!\!\!\! & \!\!\!\!\!\!\underset{\boldsymbol{w}_{i}\in\mathbb{C}^{n_{o}N_{f}}}{\mbox{min}}\left\Vert \boldsymbol{w}_{i}\right\Vert _{0}\mbox{\,\,\,\,\mbox{s.t. }\,\,\,\,\ensuremath{\left\Vert \boldsymbol{L}_{y}^{-1}(\boldsymbol{R}_{yy}\boldsymbol{w}_{i}-\boldsymbol{r}_{\Delta,i})\right\Vert _{2}^{2}}\ensuremath{\leq}}\delta_{eq,i}.\label{eq:R_yy_min_prob_cse}
\end{eqnarray}
}{\small \par}

\textcolor{black}{Note that the sparsifying dictionaries in (\ref{eq:L_y_h}),
(\ref{eq:U_telda_y}) and (\ref{eq:R_yy_min_prob_cse}) are $\boldsymbol{L}_{y}^{H}$,
$\boldsymbol{D}_{y}^{\frac{1}{2}}\boldsymbol{U}_{y}^{H}$ and $\boldsymbol{R}_{yy}$,
respectively. Furthermore, the matrix $\boldsymbol{K}$ is an identity
matrix in all cases except in (\ref{eq:R_yy_min_prob_cse}), where
it is equal to $\boldsymbol{L}_{y}^{-1}$. Additionally, some possible
sparsifying dictionaries that can be used to design a sparse FFF matrix
filter, given in (\ref{eq:MIMO-DFE-FFF}), are shown in Table \ref{tab:Examples-of-different}.
It is worth pointing out that several other sparsifying dictionaries
can be used to sparsely design FIR LEs, FBF and FFF matrix taps. In
the interest of space, we have presented above few design problems
with some possible choices for the sparsifying dictionaries and the
other choices can be derived by applying suitable transformations
to the given design problem. }
\begin{table}
\textcolor{black}{\scriptsize{}\caption{{\footnotesize{}Examples of different sparsifying dictionaries that
can be used to design $\overline{\boldsymbol{w}}_{f}$ given in (\ref{eq:MIMO-DFE-FFF})
.}\label{tab:Examples-of-different}}
}{\scriptsize \par}

\textcolor{black}{\scriptsize{}}%
\begin{tabular}[b]{|l|l|l|l|}
\hline 
\textcolor{black}{\scriptsize{} Factorization Type} & \textcolor{black}{\scriptsize{}$\boldsymbol{K}$} & \textcolor{black}{\scriptsize{}$\boldsymbol{\varPhi}$} & \textcolor{black}{\scriptsize{}$\boldsymbol{d}$ }\tabularnewline
\hline 
\hline 
\multirow{2}{*}{\textcolor{black}{\scriptsize{}$\boldsymbol{R}_{yy}=\boldsymbol{L}_{y}\boldsymbol{L}_{y}^{H}$ }} & \textcolor{black}{\scriptsize{}$\boldsymbol{I}$} & \textcolor{black}{\scriptsize{}$\boldsymbol{I}_{n_{i}}\otimes\boldsymbol{L}_{y}^{H}$} & \textcolor{black}{\scriptsize{}$\textbf{vec}(\boldsymbol{L}_{y}^{-1}\boldsymbol{\overline{\beta}})$}\tabularnewline
\cline{2-4} 
 & \textcolor{black}{\scriptsize{}$\boldsymbol{L}_{y}^{-1}$} & \textcolor{black}{\scriptsize{}$\boldsymbol{I}_{n_{i}}\otimes\boldsymbol{R}_{yy}$} & \textcolor{black}{\scriptsize{}$\textbf{vec}(\boldsymbol{\overline{\beta}})\vphantom{\boldsymbol{L}_{y}^{-H}}$}\tabularnewline
\hline 
\hline 
\textcolor{black}{\scriptsize{}$\boldsymbol{R}_{yy}=\boldsymbol{P}_{y}\boldsymbol{\Lambda}_{y}\boldsymbol{P}_{y}^{H}$} & \textcolor{black}{\scriptsize{}$\boldsymbol{I}$} & \textcolor{black}{\scriptsize{}$\boldsymbol{I}_{n_{i}}\otimes\boldsymbol{\Lambda}_{y}^{\frac{1}{2}}\boldsymbol{P}_{y}^{H}$} & \textcolor{black}{\scriptsize{}$\textbf{vec}(\boldsymbol{\Lambda}_{y}^{-\frac{1}{2}}\boldsymbol{P}_{y}^{-1}\boldsymbol{\overline{\beta}})$}\tabularnewline
\hline 
\hline 
\multirow{2}{*}{\textcolor{black}{\scriptsize{}$\boldsymbol{R}_{yy}=\boldsymbol{U}_{y}\boldsymbol{D}_{y}\boldsymbol{U}_{y}^{H}$}} & \textcolor{black}{\scriptsize{}$\boldsymbol{D}_{y}^{-\frac{1}{2}}\boldsymbol{U}_{y}^{H}$} & \textcolor{black}{\scriptsize{}$\boldsymbol{I}_{n_{i}}\otimes\boldsymbol{R}_{yy}$} & \textcolor{black}{\scriptsize{}$\textbf{vec}(\boldsymbol{\overline{\beta}})$}\tabularnewline
\cline{2-4} 
 & \textcolor{black}{\scriptsize{}$\boldsymbol{I}$} & \textcolor{black}{\scriptsize{}$\boldsymbol{I}_{n_{i}}\otimes\boldsymbol{D}_{y}^{\frac{1}{2}}\boldsymbol{U}_{y}^{H}$} & \textcolor{black}{\scriptsize{}$\textbf{vec}(\boldsymbol{D}_{y}^{-\frac{1}{2}}\boldsymbol{U}_{y}^{H}\boldsymbol{\overline{\beta}})$}\tabularnewline
\hline 
\end{tabular}{\scriptsize \par}
\end{table}
\textcolor{black}{{} }

\textcolor{black}{So far, we have shown that the problem of designing
sparse FIR filters can be cast into one of sparse approximation of
a vector by a fixed dictionary. The general form of this problem is
given by (\ref{eq:propFW}). To solve this problem, we use the well-known
Orthogonal Matching Pursuit (OMP) greedy algorithm \cite{omp07} that
estimates $\widehat{\boldsymbol{z}}_{s}$ by iteratively selecting
a set $S$ of the sparsifying dictionary columns (i.e., atoms $\boldsymbol{\phi}_{i}$'s)
of $\boldsymbol{\varPhi}$ that are most correlated with the data
vector $\boldsymbol{d}$ and then solving a restricted least-squares
problem using the selected atoms. The OMP stopping criterion  can
be either a predefined sparsity level (number of nonzero entries)
of $\boldsymbol{z_{s}}$ or  an upper-bound on the norm of the residual
error. We work with the latter case in our problem but change the
stopping criterion from an upper-bound on the norm of the residual
error to an upper-bound on the norm of ``the Projected Residual Error
(PRE)'', i.e., ``$\boldsymbol{K}\times\left(\boldsymbol{\varPhi}\boldsymbol{z}-\boldsymbol{d}\right)$''.
Note that the stopping criterion becomes a function of $\boldsymbol{K}$,
and hence this value has to be passed to the OMP algorithm to determine
$\epsilon$, i.e., $\widehat{\boldsymbol{z}}_{s}\triangleq\text{OMP}\left(\boldsymbol{\Phi},\boldsymbol{d},\boldsymbol{K},\epsilon\right)$.
}\textit{\textcolor{black}{}}\textcolor{black}{The computations involved
in the OMP algorithm are well documented in the sparse approximation
literature (e.g., \cite{omp07}) and are omitted here for the sake
of brevity. }

\textcolor{black}{Note that unlike conventional compressive sensing
techniques \cite{CS}, where the measurement matrix is a fat matrix,
the sparsifying dictionary in our framework is either a tall matrix
(fewer columns than rows) with full column rank as in (\ref{eq:ex_R_delta_1})
and (\ref{eq:ex_R_delta_2}) or a square one with full rank as in
(\ref{eq:L_y_h})--(\ref{eq:R_yy_min_prob_cse}). However, OMP and
similar methods can still be used for obtaining $\widehat{\boldsymbol{z}}_{s}$
if $\boldsymbol{R}_{yy}$ and $\boldsymbol{R}^{\perp}$ can be decomposed
into $\boldsymbol{\Psi\varPsi^{H}}$ and the data vector $\boldsymbol{d}$
is compressible \cite{sparsefeng2012,sparseFilterDesign13}.}

\textcolor{black}{Our next challenge is to determine the best sparsifying
dictionary for use in our framework. We know from the sparse approximation
literature that the sparsity of the OMP solution tends to be inversely
proportional to the worst-case coherence $\mu\left(\boldsymbol{\varPhi}\right)$,
where }\textcolor{black}{\small{}$\mu\left(\boldsymbol{\varPhi}\right)\triangleq\underset{i\neq j}{\mbox{max}}\frac{\left|\left\langle \phi_{i},\,\phi_{j}\right\rangle \right|\,}{\left\Vert \phi_{i}\right\Vert _{2}\left\Vert \phi_{j}\right\Vert _{2}}$}\textcolor{black}{{}
\cite{finiteSparseFilter013,greedIsGood03}. Notice that $\mu\left(\boldsymbol{\varPhi}\right)\in\left[0,1\right]$.
Next, we investigate the coherence of the dictionaries involved in
our setup.}

\subsection{\textcolor{black}{Worst-Case Coherence Analysis\label{subsec:Preliminary-Analysis} }}

\textcolor{black}{We carry out a coherence metric analysis to gain
some insights into the performance of different sparsifying dictionaries
and the behavior of the resulting sparse FIR filters. First and foremost,
we are concerned with analyzing $\mu\left(\boldsymbol{\varPhi}\right)$
to ensure that it does not approach $1$ for any of the proposed sparsifying
dictionaries. In addition, we are interested in identifying which
$\boldsymbol{\varPhi}$ has the smallest coherence and, hence, gives
the sparsest FIR design. While we have many sparsifying dictionaries
($\boldsymbol{R}_{yy}$, $\boldsymbol{R}^{\perp}$ and their factors)
involved in our analysis, we can classify them into two groups. The
first group is the dictionaries resulting from factorization of the
posterior error covariance matrix $\boldsymbol{R}^{\perp}$, while
the second group is either the output auto-correlation matrix $\boldsymbol{R}_{yy}$
itself or any of its factors. The matrices in the first group can
be considered asymptotically stationary Toeplitz matrices as will
be shown in Section \ref{subsec:Reduced-Complexity-Design}. In the
second group, $\boldsymbol{R}_{yy}$ is a Hermitian positive-definite
square Toeplitz (or block Toeplitz) matrix. }

\textcolor{black}{We proceed as follows to characterize the upper-bounds
on $\mu\left(\boldsymbol{\Phi}\right)$ for each kind of dictionary.
We obtain upper bounds on the worst-case coherence of both $\boldsymbol{R}^{\perp}$
and $\boldsymbol{R}_{yy}$ separately and evaluate their closeness
to $1$. Then, we demonstrate heuristically, and then through simulation,
that the coherence of the factors of $\boldsymbol{R}^{\perp}$ and
$\boldsymbol{R}_{yy}$ will be less than $1$ and smaller than that
of $\mu(\boldsymbol{R}^{\perp})$ and $\mu(\boldsymbol{R}_{yy})$,
respectively. Notice that the other dictionaries, which result from
decomposing $\boldsymbol{R}^{\perp}$ and $\boldsymbol{R}_{yy}$,
can be considered as square roots of them in the spectral-norm sense.
For example, $\left\Vert \boldsymbol{R}_{yy}\vphantom{\boldsymbol{L}^{H}}\right\Vert _{2}=\left\Vert \boldsymbol{L}_{y}\boldsymbol{L}_{y}^{H}\right\Vert _{2}\leq\left\Vert \boldsymbol{L}_{y}^{H}\boldsymbol{\vphantom{\boldsymbol{L}^{H}}}\right\Vert _{2}^{2}$
and $\left\Vert \boldsymbol{R}^{\perp}\vphantom{\boldsymbol{\boldsymbol{U}}_{\delta}^{H}}\right\Vert _{2}=\left\Vert \boldsymbol{\boldsymbol{U}}_{\perp}\boldsymbol{D}_{\perp}\boldsymbol{\boldsymbol{U}}_{\perp}^{H}\right\Vert _{2}^{2}\leq\left\Vert \boldsymbol{D}_{\perp}^{1/2}\boldsymbol{\boldsymbol{U}}_{\perp}^{H}\right\Vert _{2}^{2}$. }

\textcolor{black}{The covariance matrix $\boldsymbol{R}^{\perp}$
in (\ref{eq:MSE-MIMO-BRB}) can be expressed compactly in terms of
the SNR and CIR coefficients as $\boldsymbol{R}^{\perp}=\left[\boldsymbol{R}_{xx}^{-1}+\boldsymbol{H}^{H}\boldsymbol{R}_{nn}^{-1}\boldsymbol{H}\right]^{-1}=\left[\boldsymbol{I}+\mbox{SNR}\left(\boldsymbol{H}^{H}\boldsymbol{H}\right)\right]^{-1}$.
This shows that, at low SNR, the noise dominates, i.e., $\boldsymbol{R}^{\perp}\approx\boldsymbol{I}$,
and, consequently, $\mu\left(\boldsymbol{R}^{\perp}\right)\rightarrow0$.
As the SNR increases, the noise effect decreases and the CIR effect
starts to appear, which makes $\mu\left(\boldsymbol{R}^{\perp}\right)$
converge to a constant. Typically, this constant, as shown through
simulations, does not approach $1$. }

\textcolor{black}{On the other hand, $\boldsymbol{R}_{yy}$ has a
well-structured (Hermitian Toeplitz) closed-form in terms of the CIR
coefficients, filter time span $N_{f}$ and SNR, i.e., $\boldsymbol{R}_{yy}=\boldsymbol{H}\boldsymbol{H}^{H}+\mbox{\ensuremath{\frac{\mbox{1}}{SNR}}}\boldsymbol{I}$.
Also, it is a square matrix with full rank, due to the presence of
noise, and can be expressed in matrix form as }

\textit{\textcolor{black}{
\begin{equation}
\boldsymbol{R}_{yy}=\mbox{Toeplitz}\overbrace{\left(\left[\begin{array}{ccccccc}
r_{0} & r_{1} & \ldots & r_{v} & 0 & \ldots & 0\end{array}\right]\right)}^{\boldsymbol{\phi}_{1}^{H}}\,,\label{eq:R_yy_matrix_from}
\end{equation}
}}\textcolor{black}{where }\textcolor{black}{\small{}$r_{0}={\displaystyle \sum_{i=0}^{v}\left|h_{i}\right|^{2}+\left(\mbox{SNR}\right)^{-1}}$
and $r_{j}=\sum_{i=j}^{v}h_{i}h_{i-j}^{*},\,\forall j\neq0$. In \cite{ourFWg},
}\textcolor{black}{we showed that the worst CIR vector $\boldsymbol{h}$,
which is then used to estimate an upper-bound on $\mu(\boldsymbol{R}_{yy})$
for any given channel length $v$, can be derived by solving the following
optimization problem }

\textit{\textcolor{black}{
\begin{equation}
\widehat{\boldsymbol{h}}\triangleq\underset{\boldsymbol{h}}{\mbox{\mbox{arg}\mbox{max}}}\,\left|\boldsymbol{h}^{H}\boldsymbol{R}\boldsymbol{h}\right|\,\,\,\mbox{subject to}\,\,\,\boldsymbol{h}^{H}\boldsymbol{h}=1\,,\label{eq:quadEqProb}
\end{equation}
}}\textcolor{black}{where $\boldsymbol{h}=\left[\begin{array}{cccc}
h_{0} & h_{1} & \ldots & h_{v}\end{array}\right]^{H}$ is the length-$(v+1)$ CIR vector and $\boldsymbol{R}$ is a matrix
that has ones along the super and sub-diagonals. It is known that
the solution of (\ref{eq:quadEqProb}) is the eigenvector corresponding
to the maximum eigenvalue of $\boldsymbol{R}$. The eigenvalues $\lambda_{s}$
and eigenvectors $h_{j}^{(s)}$ of the matrix $\boldsymbol{R}$ have
the following simple closed-forms \cite{eigValueVector_R}}

\textit{\textcolor{black}{
\begin{eqnarray}
\lambda_{s} & = & 2\,\mbox{cos}\left(\frac{\pi s}{v+2}\right)\,,\,h_{j}^{(s)}=\sqrt{\frac{2}{v+2}}\mbox{sin\ensuremath{\left(\frac{j\pi s}{v+2}\right),}}\label{eq:worst-taps}
\end{eqnarray}
}}\textcolor{black}{where $s,j=1,\ldots,v+1.$ By numerically evaluating
$h_{j}^{(s)}$ for the maximum $\left|\lambda_{s}\right|$, we find
that the worst-case coherence of $\boldsymbol{R}_{yy}$ (for any $v$)
is sufficiently less than $1$. This observation points to the likely
success of OMP in providing the sparsest solution $\widehat{\boldsymbol{z}}_{s}$
which corresponds to the dictionary that has the smallest $\mu(\boldsymbol{R}_{yy})$.
Next, we propose a novel approach to perform the involved matrix factorizations
in a reduced-complexity fashion. }

\subsection{\textcolor{black}{Reduced-Complexity Design \label{subsec:Reduced-Complexity-Design}}}

\textcolor{black}{In this section, we propose reduced-complexity designs
for the FIR filters discussed above, including LEs and DFEs, for MIMO
systems. The proposed designs in Section \ref{sec:Sparse-FIR-Equalization}
involve Cholesky factorization and/or eigen decomposition, whose computational
costs could be large for channels with large delay spreads. For a
Toeplitz matrix, the most efficient algorithms for Cholesky factorization
are Levinson or Schur algorithms \cite{statDSP}, which involve $\mathcal{O}(M^{2})$
computations, where $M$ is the matrix dimension. In contrast, since
a circulant matrix is asymptotically equivalent to a Toeplitz matrix
for reasonably large dimension \cite{toep2circApp2003}, the eigen
decomposition of a circulant matrix can be computed efficiently using
the fast Fourier transform (FFT) and its inverse with only $\mathcal{O}\left(M\mbox{log}_{2}(M)\right)$
operations}\footnote{\textcolor{black}{Toeplitz and circulant matrices are asymptotic in
the output block length which is equal to the time span (not number
of nonzero taps) of the FFF. This asymptotic equivalence implies that
the eigenvalues of the two matrices behave similarly. Furthermore,
it also implies that factors, products, and inverses behave similarly
\cite{gray1972asymptotic}.}}\textcolor{black}{. We can use this asymptotic equivalence between
Toeplitz and circulant matrices to carry out the computations needed
for $\boldsymbol{R}_{yy},$ and $\boldsymbol{R}^{\perp}$ factorizations
efficiently using the FFT and inverse FFT. In addition, direct matrix
inversion can be avoided when computing the coefficients of the filters.
This approximation turns out to be quite accurate from simulations
as will be shown later. }

\textcolor{black}{It is well known that a circulant matrix, $\boldsymbol{C}$,
has the discrete Fourier transform (DFT) basis vectors as its eigenvectors
and the DFT of its first column as its eigenvalues. Thus, an $M\times M$
circulant matrix $\boldsymbol{C}$ can be decomposed as $\boldsymbol{C}$
=$\frac{1}{M}\boldsymbol{F}_{M}^{H}\boldsymbol{\varLambda}_{\boldsymbol{c}}\boldsymbol{F}_{M}$}\textcolor{black}{\small{},}\textcolor{black}{{}
where $\boldsymbol{F}_{M}$ is the DFT matrix with $f_{k,l}=e^{-j2\pi kl/M}$,
$0\leq k,\,l\leq M-1$, and $\boldsymbol{\varLambda}_{\boldsymbol{c}}$
is an $M\times M$ diagonal matrix whose diagonal elements are the
$M$-point DFT of $\boldsymbol{c}=\left\{ c\right\} _{i=0}^{i=M-1}$,
the first column of the circulant matrix. Further, from the orthogonality
of DFT basis functions, $\boldsymbol{F}_{M}^{H}\boldsymbol{F}_{M}=\boldsymbol{F}_{M}\boldsymbol{F}_{M}^{H}=M\,\boldsymbol{I}_{M}$
and $\boldsymbol{F}_{N}^{H}\boldsymbol{F}_{N}=M\,\boldsymbol{I}_{N+1}$
where $\boldsymbol{F}_{N}$ is an $M\times N$ matrix, but $\boldsymbol{F}_{N}\boldsymbol{F}_{N}^{H}\neq M\,\boldsymbol{I}_{N+1}$
and instead $\boldsymbol{F}_{N}\boldsymbol{F}_{N}^{H}=N\left[\begin{array}{ccc}
\boldsymbol{I}_{N} & \ldots & \boldsymbol{I}_{N}\end{array}\right]^{T}\left[\begin{array}{ccc}
\boldsymbol{I}_{N} & \ldots & \boldsymbol{I}_{N}\end{array}\right]$.}

\textcolor{black}{We denote by $\overline{\boldsymbol{R}}_{yy},\,\overline{\boldsymbol{R}}_{yx}$,
and $\overline{\boldsymbol{R}}^{\perp}$ the circulant approximations
to the matrices $\boldsymbol{R}_{yy},\,\boldsymbol{R}_{yx}\mbox{,}$
$\mbox{ and }\boldsymbol{R}^{\perp}$ respectively. In addition, we
denote the noiseless channel output vector as $\widetilde{\boldsymbol{y}},$
i.e., $\widetilde{\boldsymbol{y}}=\boldsymbol{Hx}$. We first derive
the circulant approximation for the block Toeplitz matrix $\boldsymbol{R}_{yy}$
when $n_{o}\geq2$, and the case of SISO systems follows as a special
case of the block Toeplitz case by setting $n_{o}=n_{i}=1$. }

\textcolor{black}{The autocorrelation matrix $\boldsymbol{\overline{R}}_{yy}$
is computed as }

\textit{\textcolor{black}{
\begin{equation}
\boldsymbol{\overline{R}}_{yy}=\underbrace{E\left[\widetilde{\boldsymbol{y}}_{k}\widetilde{\boldsymbol{y}}_{k}\right]}_{\overline{\boldsymbol{R}}_{\widetilde{y}\widetilde{y}}}+\underbrace{\frac{1}{SNR}}_{\sigma_{n}^{2}}\boldsymbol{I}_{N_{f}}.
\end{equation}
}}\textcolor{black}{{} To approximate the block Toeplitz $\boldsymbol{R}_{yy}$
as a circulant matrix, we assume that $\left\{ \widetilde{\boldsymbol{y}}_{k}\right\} $
is cyclic. Hence, $E\left[\widetilde{\boldsymbol{y}}_{k}\widetilde{\boldsymbol{y}}_{k}\right]$
can be approximated as a time-averaged autocorrelation function as
follows (defining $L=n_{o}N_{f}$)}

\textit{\textcolor{black}{
\begin{eqnarray}
\!\!\!\!\!\overline{\boldsymbol{R}}_{\widetilde{y}\widetilde{y}} & \!\!\!\!= & \!\!\!\frac{1}{N_{f}}\sum_{k=0}^{N_{f}-1}\widetilde{\boldsymbol{y}}_{k}\widetilde{\boldsymbol{y}}_{k}^{H}=\frac{1}{N_{f}}\boldsymbol{C}_{\underline{Y}}\boldsymbol{C}_{\underline{Y}}^{H}\nonumber \\
\!\!\!\! & \!\!\!\!= & \!\!\!\frac{1}{N_{f}}\left(\frac{1}{L}\boldsymbol{F}_{L}^{H}\boldsymbol{\varLambda}_{\underline{\widetilde{Y}}}\boldsymbol{F}_{N_{f}}\right)\left(\frac{1}{L}\boldsymbol{F}_{N_{f}}^{H}\boldsymbol{\varLambda}_{\underline{\widetilde{Y}}}^{H}\boldsymbol{F}_{L}\right)\nonumber \\
 & = & \!\!\!\frac{1}{L^{2}}\boldsymbol{F}_{L}^{H}\boldsymbol{\varLambda}_{\widetilde{\underline{Y}}}\left[\begin{array}{c}
\boldsymbol{I}_{N_{f}}\\
\vdots\\
\boldsymbol{I}_{N_{f}}
\end{array}\right]\underbrace{\left[\begin{array}{ccc}
\boldsymbol{I}_{N_{f}} & \ldots & \boldsymbol{I}_{N_{f}}\end{array}\right]}_{n_{o}\,\mbox{blocks}}\boldsymbol{\varLambda}_{\underline{\widetilde{Y}}^{H}}\boldsymbol{F}_{L}\nonumber \\
 & = & \!\!\!\frac{1}{L^{2}}\boldsymbol{F}_{L}^{H}\!\left[\!\!\begin{array}{c}
\boldsymbol{\varLambda}_{\underline{\widetilde{Y}}^{1}}\\
\vdots\\
\boldsymbol{\varLambda}_{\widetilde{\underline{Y}}^{n_{o}}}
\end{array}\!\!\right]\!\left[\!\!\begin{array}{ccc}
\boldsymbol{\varLambda}_{\widetilde{\underline{Y}}^{1}}^{H} & \ldots & \boldsymbol{\varLambda}_{\widetilde{\underline{Y}}^{n_{o}}}^{H}\end{array}\!\!\right]\boldsymbol{F}_{L}\,,\label{eq:Ryy_circ}
\end{eqnarray}
}}\textcolor{black}{where $\boldsymbol{F}_{L}$ is a DFT matrix of
size $L\times L$, $\boldsymbol{F}_{N_{f}}$ is a DFT matrix of size
$L\times N_{f}$, the column vector $\widetilde{\boldsymbol{\underline{Y}}}$
is the $L$-point DFT of $\widetilde{\boldsymbol{y}}_{1}=\left[\begin{array}{ccc}
\widetilde{\boldsymbol{y}}_{N_{f}-1}^{T} & \widetilde{\boldsymbol{y}}_{N_{f}-2}^{T}\ldots & \widetilde{\boldsymbol{y}}_{0}^{T}\end{array}\right]$, $\widetilde{\boldsymbol{\underline{Y}}}^{i}$ is the $i^{th}$ subvector
of $\widetilde{\boldsymbol{\underline{Y}}}$, i.e., $\widetilde{\underline{\boldsymbol{Y}}}$=$\left[\begin{array}{cccc}
\widetilde{\boldsymbol{\underline{Y}}}^{1} & \widetilde{\boldsymbol{\underline{Y}}}^{2} & \ldots & \widetilde{\boldsymbol{\underline{Y}}}^{n_{o}}\end{array}\right]^{T}$, $\widetilde{\boldsymbol{y}}_{i}$ is the $n_{o}\times1$ output
vector and $\boldsymbol{C}_{y}=\mbox{\textbf{circ}(\ensuremath{\widetilde{\boldsymbol{y}}_{1}})}$
where }\textbf{\textcolor{black}{circ}}\textcolor{black}{{} denotes
a circulant matrix whose first column is $\widetilde{\boldsymbol{y}}_{1}.$
Then, }

\textit{\textcolor{black}{
\begin{eqnarray}
\!\!\boldsymbol{\overline{R}}_{yy} & \!\!\!\!=\!\!\!\! & \overline{\boldsymbol{R}}_{\widetilde{y}\widetilde{y}}+n_{o}\sigma_{n}^{2}\boldsymbol{I}_{n_{o}N_{f}}\nonumber \\
\!\!\!\! & \!\!\!\!=\!\!\!\! & \frac{1}{L^{2}}\boldsymbol{F}_{L}^{H}\!\left[\!\!\begin{array}{c}
\boldsymbol{\varLambda}_{\widetilde{\underline{Y}}^{1}}\\
\vdots\\
\boldsymbol{\varLambda}_{\widetilde{\underline{Y}}^{n_{o}}}
\end{array}\!\!\right]\underbrace{\!\!\left[\!\!\begin{array}{ccc}
\boldsymbol{\varLambda}_{\widetilde{\underline{Y}}^{1}}^{H} & \ldots & \boldsymbol{\varLambda}_{\widetilde{\underline{Y}}^{n_{o}}}^{H}\end{array}\!\!\right]\!\!}_{\boldsymbol{\Psi}_{\underline{Y}}^{H}}\boldsymbol{F}_{L}+n_{o}\sigma_{n}^{2}\boldsymbol{I}_{L}\nonumber \\
\!\!\!\! & \!\!\!\!=\!\!\!\! & \frac{1}{L^{2}}\boldsymbol{F}_{L}^{H}\left(\boldsymbol{\Psi}_{\underline{Y}}\boldsymbol{\Psi}_{\underline{Y}}^{H}+n_{o}L\sigma_{n}^{2}\boldsymbol{I}_{L}\right)\boldsymbol{F}_{L}=\boldsymbol{\varSigma\varSigma}^{H}.\label{eq:R_yy_circ}
\end{eqnarray}
}}\textcolor{black}{{} Using the matrix inversion lemma \cite{matAnalysis},
the inverse of $\boldsymbol{\overline{R}}_{yy}$ is }

\textit{\textcolor{black}{
\begin{eqnarray}
\!\!\boldsymbol{\overline{R}}_{yy}^{-1} & \!\!\!=\!\!\! & \left\{ \frac{1}{L^{2}}\boldsymbol{F}_{L}^{H}\left(\boldsymbol{\Psi}_{\underline{Y}}\boldsymbol{\Psi}_{\underline{Y}}^{H}+n_{o}L\sigma_{n}^{2}\boldsymbol{I}_{L}\right)\boldsymbol{F}_{L}\right\} ^{-1}\nonumber \\
 & \!\!\!=\!\!\! & \boldsymbol{F}_{L}^{H}\left(\boldsymbol{\Psi}_{\underline{Y}}\boldsymbol{\Psi}_{\underline{Y}}^{H}+n_{o}L\sigma_{n}^{2}\boldsymbol{I}_{L}\right)^{-1}\boldsymbol{F}_{L}\nonumber \\
 & \!\!\!=\!\!\! & \frac{1}{n_{o}L\sigma_{n}^{2}}\boldsymbol{F}_{L}^{H}\left(\boldsymbol{I}_{L}-\boldsymbol{\Psi}_{\underline{Y}}\boldsymbol{\varLambda}_{\varrho}^{-1}\boldsymbol{\Psi}_{\underline{Y}}^{H}\right)\boldsymbol{F}_{L}.\label{eq:R_yy_inv_circ-mimo_le}
\end{eqnarray}
}}\textcolor{black}{where }\textit{\textcolor{black}{$\varrho=\underbrace{\sum_{i=1}^{n_{o}}\left|\left\Vert \widetilde{\boldsymbol{\underline{Y}}}^{i}\right\Vert \right|^{2}}_{\overline{\varrho}}+n_{o}L\sigma_{n}^{2}\boldsymbol{1}_{L}$}}\textcolor{black}{.
Here, $\left|\left\Vert .\vphantom{\widetilde{\boldsymbol{\underline{Y}}}}\right\Vert \right|^{2}$
is defined as the element-wise norm square}

\textit{\textcolor{black}{
\begin{equation}
\left|\left\Vert \left[\begin{array}{ccc}
a_{0} & \!\!\ldots & \!a_{N_{f}-1}\end{array}\right]^{H}\right\Vert \right|^{2}\!\!=\!\!\left[\!\!\begin{array}{ccc}
\left|a_{0}\right|^{2}\!\! & \!\ldots & \!\!\left|a_{N_{f}-1}\right|^{2}\end{array}\!\!\right]^{H}.
\end{equation}
}}

\textcolor{black}{Notice that }\textcolor{black}{\small{}$\boldsymbol{\Psi}_{\underline{Y}}\boldsymbol{\Psi}_{\underline{Y}}^{H}=\sum_{i=1}^{n_{o}}\left|\left\Vert \boldsymbol{\widetilde{\underline{Y}}}^{i}\right\Vert \right|^{2}=N_{f}\sum_{i=1}^{n_{o}}\left|\left\Vert \boldsymbol{\underline{H}}^{i}\right\Vert \right|^{2}.$}\textcolor{black}{{}
Without loss of generality, we can write the noiseless channel output
sequence $\widetilde{\boldsymbol{y}}_{k}$ in the discrete frequency
domain as a column vector as follows}

\textcolor{black}{\small{}
\begin{eqnarray}
\widetilde{\boldsymbol{\underline{Y}}} & = & \boldsymbol{\underline{H}}^{H}\odot\underline{\boldsymbol{P}}_{\Delta}\odot\widetilde{\boldsymbol{\underline{X}}}
\end{eqnarray}
}\textcolor{black}{where $\odot$ denotes element-wise multiplication,
}\textcolor{black}{\small{}$\widetilde{\boldsymbol{\underline{X}}}=\left[\begin{array}{ccc}
\boldsymbol{\underline{X}}^{T} & \ldots & \boldsymbol{\underline{X}}^{T}\end{array}\right]^{T}$ where $\boldsymbol{\underline{X}}$}\textcolor{black}{{} is the DFT
of the data vector, }\textcolor{black}{\small{}$\boldsymbol{\underline{P}}_{\Delta}=\left[\begin{array}{ccc}
\widetilde{\boldsymbol{\underline{P}}}_{\Delta}^{T} & \ldots & \widetilde{\boldsymbol{\underline{P}}}_{\Delta}^{T}\end{array}\right]^{T}$, $\widetilde{\boldsymbol{\underline{P}}}_{\Delta}=\left[\begin{array}{cccc}
1 & e^{-j2\pi\Delta/N_{f}} & \ldots & e^{-j2\pi\left(N_{f}-1\right)\Delta/N_{f}}\end{array}\right]^{T}$, }\textcolor{black}{and $\boldsymbol{\underline{H}}$ is the DFT
of the CIRs, }\textcolor{black}{\small{}$\boldsymbol{\underline{H}}=\left[\begin{array}{ccc}
\boldsymbol{\underline{H}}^{1T} & \ldots & \boldsymbol{\underline{H}}^{n_{o}T}\end{array}\right]^{T}$}\textcolor{black}{. To illustrate, for $n_{o}=1$, $\boldsymbol{\overline{R}}_{yy}$
in (\ref{eq:R_yy_circ}) reduces to }

\textcolor{black}{\small{}
\begin{eqnarray}
\boldsymbol{\overline{R}}_{yy} & = & \overline{\boldsymbol{R}}_{\widetilde{y}\widetilde{y}}+\sigma_{n}^{2}\boldsymbol{I}_{N_{f}}=\boldsymbol{F}_{N_{f}}^{H}(\boldsymbol{\varLambda}_{\varrho_{1}})\boldsymbol{F}_{N_{f}}=\boldsymbol{Q}\boldsymbol{Q}^{H}\,,\label{eq:R_yy_circ-1}
\end{eqnarray}
}\textcolor{black}{where $\varrho_{1}=N_{f}\left|\left\Vert \boldsymbol{\underline{H}}\right\Vert \right|^{2}+\sigma_{n}^{2}N_{f}\boldsymbol{1}_{N_{f}}$,
$\boldsymbol{\underline{H}}$ is the $N_{f}$-point DFT of the CIR
$\boldsymbol{h}$ and $\boldsymbol{\underline{P}}_{\Delta}=\widetilde{\boldsymbol{\underline{P}}}_{\Delta}$.
Similarly, after some algebraic manipulations, $\overline{\boldsymbol{R}}^{\perp}$
can be expressed as}

\textit{\textcolor{black}{
\begin{eqnarray}
\overline{\boldsymbol{R}}^{\perp} & = & \frac{1}{L}\boldsymbol{F}_{N}^{H}\!\!\left(\!\!\boldsymbol{I}_{N}-\!\!\left[\!\!\begin{array}{c}
\boldsymbol{I}_{M}\\
\vdots\\
\boldsymbol{I}_{M}
\end{array}\right]\!\boldsymbol{\varLambda}_{\overline{\varrho}\varoslash\varrho}\!\left[\begin{array}{ccc}
\boldsymbol{I}_{M} & \ldots & \boldsymbol{I}_{M}\end{array}\!\right]\!\!\right)\boldsymbol{F}_{N}\nonumber \\
 & = & \boldsymbol{\varTheta}\boldsymbol{\varTheta}^{H}\,,\label{eq:r_delta_perpend}
\end{eqnarray}
}}\textcolor{black}{where $\varoslash$ denotes element-wise division
and $N=n_{i}(N_{f}+v)$. Notice that in the special case of SISO systems,
i.e., $n_{i}=n_{o}=1$, $\overline{\boldsymbol{R}}^{\perp}$ can be
expressed as follows}

\textit{\textcolor{black}{
\begin{eqnarray}
\overline{\boldsymbol{R}}^{\perp} & = & \boldsymbol{R}_{xx}-\boldsymbol{\overline{R}}_{yx}^{H}\boldsymbol{\overline{R}}_{yy}^{-1}\boldsymbol{\overline{R}}_{yx}\nonumber \\
 & = & \boldsymbol{I}_{N}-\boldsymbol{\overline{R}}_{xy}\left\{ \frac{1}{N\sigma_{n}^{2}}\boldsymbol{F}_{N}^{H}\left(\boldsymbol{\varLambda}_{\widetilde{\underline{Y}}}\boldsymbol{\varLambda}_{\theta}^{-1}\boldsymbol{\varLambda}_{\widetilde{\underline{X}}}^{H}\right)\boldsymbol{F}_{N}\right\} \nonumber \\
 & = & \boldsymbol{I}_{N}-\left\{ \frac{1}{N^{2}}\boldsymbol{F}_{N}^{H}\left(\boldsymbol{\varLambda}_{\underline{X}}\boldsymbol{\varLambda}_{\widetilde{\underline{Y}}}^{H}\boldsymbol{\varLambda}_{\widetilde{\underline{Y}}}\boldsymbol{\varLambda}_{\theta}^{-1}\boldsymbol{\varLambda}_{\underline{X}}^{H}\right)\boldsymbol{F}_{N}\right\} \nonumber \\
 & = & \frac{1}{N^{2}}\boldsymbol{F}_{N}^{H}\left(N\,\boldsymbol{I}_{N}-\boldsymbol{\varLambda}_{\underline{X}}\boldsymbol{\varLambda}_{\left(\overline{\theta}\varoslash\theta\right)}\boldsymbol{\varLambda}_{\underline{X}}^{H}\right)\boldsymbol{F}_{N}\nonumber \\
 & = & \frac{1}{N}\boldsymbol{F}_{N}^{H}\left(\boldsymbol{I}_{N}-\boldsymbol{\varLambda}_{\left(\overline{\theta}\varoslash\theta\right)}\right)\boldsymbol{F}_{N}=\boldsymbol{\varGamma}\boldsymbol{\varGamma}^{H},\label{eq:R_delta_circ}
\end{eqnarray}
}}\textcolor{black}{where $N=N_{f}+v$, $\boldsymbol{F}_{N}$ is an
$N\times N$ DFT matrix, $\boldsymbol{F}_{N_{f}}$ is an $N\times N_{f}$
DFT matrix, $\theta=\overline{\theta}+N\sigma_{n}^{2}\boldsymbol{1}_{N}$
and $\overline{\theta}=\left|\left\Vert \widetilde{\underline{\boldsymbol{Y}}}\right\Vert \right|^{2}$.
Note that $\widetilde{\underline{\boldsymbol{Y}}}$ is the $N$-point
DFT of $\left[\begin{array}{cccc}
\widetilde{y}_{N_{f}}^{T} & \widetilde{y}_{N_{f}-1}^{T} & \ldots & \widetilde{y}_{1}^{T}\end{array}\right]$. }

\textcolor{black}{In summary, the proposed design method for the sparse
FIR filters involves the following steps:}
\begin{enumerate}
\item \textcolor{black}{An estimate for the channel between the input(s)
and the output(s) of the actual transmission channel is obtained.
Then, the matrices defined in Table \ref{tab:Channel-Equalization-notation}
are computed. }
\item \textcolor{black}{The required matrices involved in our design, i.e.,
$\ensuremath{\boldsymbol{R}^{\perp}}$or $\ensuremath{\boldsymbol{R}}_{yy}$,
are factorized using reduced-complexity design discussed above in
this section. }
\item \textcolor{black}{Based on a desired performance-complexity tradeoff,
$\epsilon$ is computed. Afterward, the dictionary with the smallest
coherence is selected for use in designing the sparse FIR filter.}
\item \textcolor{black}{The parameters $\boldsymbol{\varPhi}$, $\boldsymbol{d}$,
and $\boldsymbol{K}$ are jointly used to estimate the locations and
weights of the filter taps using the OMP algorithm.}
\end{enumerate}
\textcolor{black}{We conclude this section by noting that using this
low-complexity fast computation matrix factorization approach, we
are able to design the FIR filters in a reduced-complexity manner
where neither a Cholesky nor an eigen factorization is needed. Furthermore,
direct inversion of the matrices involved in the design of filters
is avoided. }

\subsection{\textcolor{black}{Complexity Analysis}}

\textcolor{black}{In this section, we evaluate the computation complexity
of various filter designs in terms of complex multiplications/additions
(CM/A). For the proposed sparse MIMO LEs and DFEs, the main computational
tasks are factorizations of the matrices $\boldsymbol{R}_{yy}$, $\boldsymbol{R}^{\perp}$
and the OMP computations. It is noted in \cite{omp07} that the computation
cost, CM/A, of OMP is $\mathcal{O}\left(MNS\right)$, where $MN$
is the size of the equalizer vector/matrix and $S$ is the number
of nonzero entries of $\boldsymbol{z}_{s}$. Note that an additional
$\mathcal{O}\left(S^{3}\right)$ CM/A computations are required to
obtain the restricted least squares estimate of $\boldsymbol{z}_{s}$
\cite{finiteSparseFilter013}. Hence, the total cost to estimate $\boldsymbol{z}_{s}$
using our proposed design method is the sum of the factiorization
cost of the involved matrices in the FIR filter design, the OMP cost,
and the restricted least squares cost, i.e., $\mathcal{O}\left(M\,\mbox{log}\left(M\right)+MNS+S^{3}\right)$,
which is typically much lower than the computational complexity of
$\mathcal{O}\left(M^{3}+NM^{2}\right)$ for convex-optimization-based
approaches \cite{finiteSparseFilter013}. Furthermore, the cost of
our proposed method is much smaller than the cost required to estimate
the optimum FIR equalizers given in \cite{mimoDFE}. The complexity
of our proposed design method as compared to the optimum equalizers
and some other sparse designs from the literature is summarized in
Table \ref{tab:-Computational-complexity}.}\textit{\textcolor{black}{{}
}}\textcolor{black}{Next, we will report the results of our numerical
experiments to evaluate the performance of our proposed framework
considering different FIR filter designs and using different sparsifying
dictionaries for each design. }

\textcolor{black}{}
\begin{table}
\textcolor{black}{\caption{\textcolor{black}{{} }\textcolor{black}{\footnotesize{}Computational
complexity of various equalizer designs.}\textcolor{black}{\label{tab:-Computational-complexity} }}
}

\textcolor{black}{\footnotesize{}}%
\begin{tabular}{|>{\raggedright}p{15ex}|>{\raggedright}p{50ex}|}
\hline 
\textcolor{black}{\scriptsize{}Equalizer Type} & \textcolor{black}{\scriptsize{}Design Complexity}\tabularnewline
\hline 
\hline 
\textcolor{black}{\scriptsize{}Optimum FIR LEs \cite{mimoDFE}} & \textcolor{black}{\scriptsize{}$\mathcal{O}\left(n_{i}\left(N_{f}+v\right)\left(n_{o}N_{f}\right)^{2}+n_{o}^{3}N_{f}^{2}\right)$}\tabularnewline
\hline 
\hline 
\multirow{2}{15ex}{\textcolor{black}{\scriptsize{}Optimum FIR DFEs \cite{mimoDFE}}} & \textcolor{black}{\scriptsize{}FBF: $\mathcal{O}\left(n_{i}^{3}\left(N_{b}+1\right)^{3}+n_{i}^{3}+n_{i}^{3}\left(N_{b}+1\right)\right)$}\tabularnewline
\cline{2-2} 
 & \textcolor{black}{\scriptsize{}FFF: $\mathcal{O}\left(n_{i}\left(N_{f}+v\right)\left(n_{o}N_{f}\right)^{2}+n_{i}^{2}\left(N_{b}+1\right)^{2}\right)$}\tabularnewline
\hline 
\hline 
\textcolor{black}{\scriptsize{}Sparse FIR LEs \cite{newDFW}} & \textcolor{black}{\scriptsize{}$\mathcal{O}\left(\left(n_{o}N_{f}\right)^{2}+n_{o}n_{i}N_{f}S+S^{3}\right)$}\tabularnewline
\hline 
\hline 
\multirow{2}{15ex}{\textcolor{black}{\scriptsize{}Sparse FIR DFEs  \cite{newDFW}}} & \textcolor{black}{\scriptsize{}FBF: $\mathcal{O}\left(\left(N_{f}+v\right)^{2}+n_{i}^{2}\left(N_{f}+v\right)^{2}S+S^{3}\right)$}\tabularnewline
\cline{2-2} 
 & \textcolor{black}{\scriptsize{}FFF: $\mathcal{O}\left(\left(n_{o}N_{f}\right)^{2}+\left(n_{o}N_{f}\right)^{2}S+S^{3}\right)$}\tabularnewline
\hline 
\hline 
\textcolor{black}{\scriptsize{}Proposed Sparse FIR LEs } & \textcolor{black}{\scriptsize{}$\mathcal{O}\left(\left(n_{o}N_{f}\right)\mbox{log}\left(n_{o}N_{f}\right)+n_{o}n_{i}N_{f}S+S^{3}\right)$}\tabularnewline
\hline 
\hline 
\multirow{2}{15ex}{\textcolor{black}{\scriptsize{}Proposed Sparse FIR DFEs}} & \textcolor{black}{\scriptsize{}FBF:$\mathcal{O}\left(\left(N_{f}+v\right)\mbox{log}\left(N_{f}+v\right)+n_{i}^{2}\left(N_{f}+v\right)^{2}S+S{}^{3}\right)$}\tabularnewline
\cline{2-2} 
 & \textcolor{black}{\scriptsize{}FFF: $\mathcal{O}\left(\left(n_{o}N_{f}\right)\mbox{log}\left(n_{o}N_{f}\right)+\left(n_{o}N_{f}\right)^{2}S+S^{3}\right)$}\tabularnewline
\hline 
\end{tabular}{\footnotesize \par}
\end{table}

\section{\textcolor{black}{Numerical Results \label{sec:Simulation-Results}}}

\textcolor{black}{We now investigate the performance of our proposed
framework. The CIRs used in our numerical results are unit-energy
symbol-spaced FIR filters with $v$ taps generated as zero-mean unit-variance
uncorrelated complex Gaussian random variables. The CIR taps are assumed
to have a uniform power-delay-profile}\footnote{\textcolor{black}{This type of CIRs can be considered as a wrost-case
assumption since the inherent sparsity of other channel models, e.g.,
\cite{ITU-A} and \cite{TG3C}, can lead to further reduce the number
of equalizer taps (i.e., sparser equalizers).}}\textcolor{black}{{} (UPDP)}\textit{\textcolor{black}{. }}\textcolor{black}{Note
that this type of channel is rather difficult to equalize because
its PDP is uniform and non-sparse. The performance results are calculated
by averaging over 5000 channel realizations. Error bars, when used,
show the confidence intervals of the data, i.e., the standard deviation
along a curve. We use the notation $\boldsymbol{D}(\boldsymbol{\chi}_{f})$
to refer to a LE designed using the sparsifying dictionary $\boldsymbol{\chi}_{f}$,
while $\boldsymbol{D}(\boldsymbol{\chi}_{b},\,\boldsymbol{\chi}_{f})$
is used to refer to a FBF designed using the sparsifying dictionary
$\boldsymbol{\chi}_{b}$ and a FFF designed using the sparsifying
dictionary $\boldsymbol{\chi}_{f}$. Note that \cite{newDFW} follows
as a special case of our proposed design method by choosing the classical
Cholesky (of the form $\boldsymbol{L}\boldsymbol{L}^{H}$) as the
factorization method and keeping the parameter $\boldsymbol{K}$ in
(\ref{eq:propFW}) always equal to the identity matrix, e.g., $\boldsymbol{K}=\boldsymbol{I}$,
$\boldsymbol{\Phi}=\boldsymbol{L}^{H}$ and $\boldsymbol{d}$ = $\boldsymbol{L}_{y}^{-1}\boldsymbol{r}_{\Delta}$.
Hence, in the results below, we have implicitly compared with the
approach proposed in \cite{newDFW} when such setting is used.}

\begin{figure}
\textcolor{black}{\includegraphics[scale=0.44]{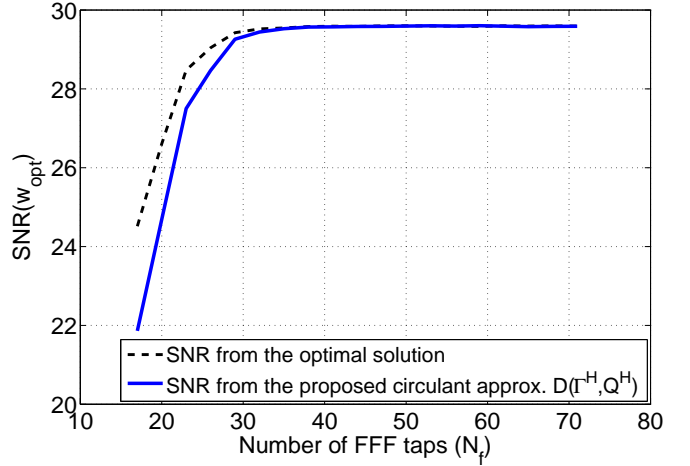}}

\textcolor{black}{\caption{{\footnotesize{}Performance of circulant approximation based approach
for UPDP channel with $v=8$ and input $\mbox{SNR}=30\,\mbox{dB}$.}\label{fig:snr_versus_nf}}
}
\end{figure}

\textcolor{black}{To quantify the accuracy of approximating Toeplitz
matrices, e.g., $\boldsymbol{R}_{yy}$ and $\boldsymbol{R}^{\perp}$,
by their equivalent circulant matrices, e.g, $\boldsymbol{\overline{R}}_{yy}$
and $\overline{\boldsymbol{R}}^{\perp}$, respectively, we plot the
optimal output SNR and the output SNR obtained from the circulant
approximation versus the number of FFF taps ($N_{f}$) in Figure \ref{fig:snr_versus_nf}.
The gap between the optimal output SNR and the output SNR from the
circulant approximation approaches zero as the number of the FFF taps
increases, as expected. A good rule of the thumb for $N_{f}$, to
obtain an accurate approximation, would be $N_{f}\geq4v$. }
\textcolor{black}{}
\begin{figure}[t]
\textcolor{black}{\includegraphics[scale=0.43]{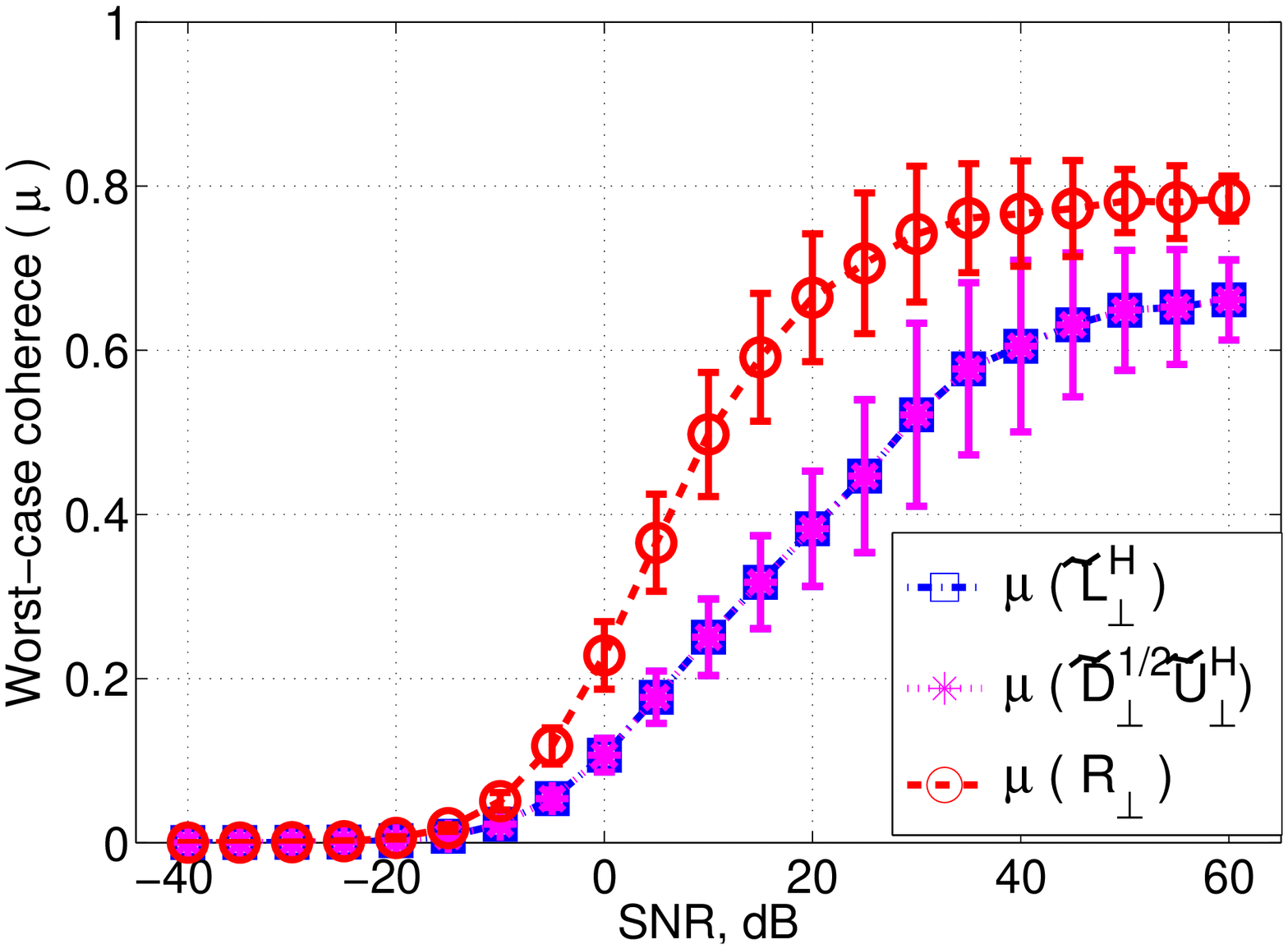}}

\textcolor{black}{\caption{{\footnotesize{}Worst-case coherence for sparsifying dictionaries
$\widetilde{\boldsymbol{L}}_{\perp}$ and $\widetilde{\boldsymbol{D}}_{\perp}^{\frac{1}{2}}\widetilde{\boldsymbol{U}}_{\perp}^{H}$
versus input SNR for UPDP with $v=8$ and $N_{f}=80$.}\textcolor{black}{\footnotesize{}
Note that we estimate $\mu\left(\widetilde{\boldsymbol{D}}_{\perp}^{\frac{1}{2}}\widetilde{\boldsymbol{U}}_{\perp}^{H}\right)$
and $\mu\left(\widetilde{\boldsymbol{L}}_{\perp}^{H}\right)$ after
removing the $\left(\Delta+1\right)^{th}$ column as discussed in
(\ref{eq:mimoDFE-FBF}).}\textcolor{black}{{} Moreover, changing the
}\textcolor{black}{\footnotesize{}$\left(\Delta+1\right)^{th}$}\textcolor{black}{{}
location has insignificant effect on }\textcolor{black}{\footnotesize{}$\mu(\boldsymbol{\varPhi})$
to show.}\textcolor{black}{{} \label{fig:coherene }}}
}
\end{figure}
\textcolor{black}{}
\begin{figure}[t]
\textcolor{black}{\includegraphics[scale=0.43]{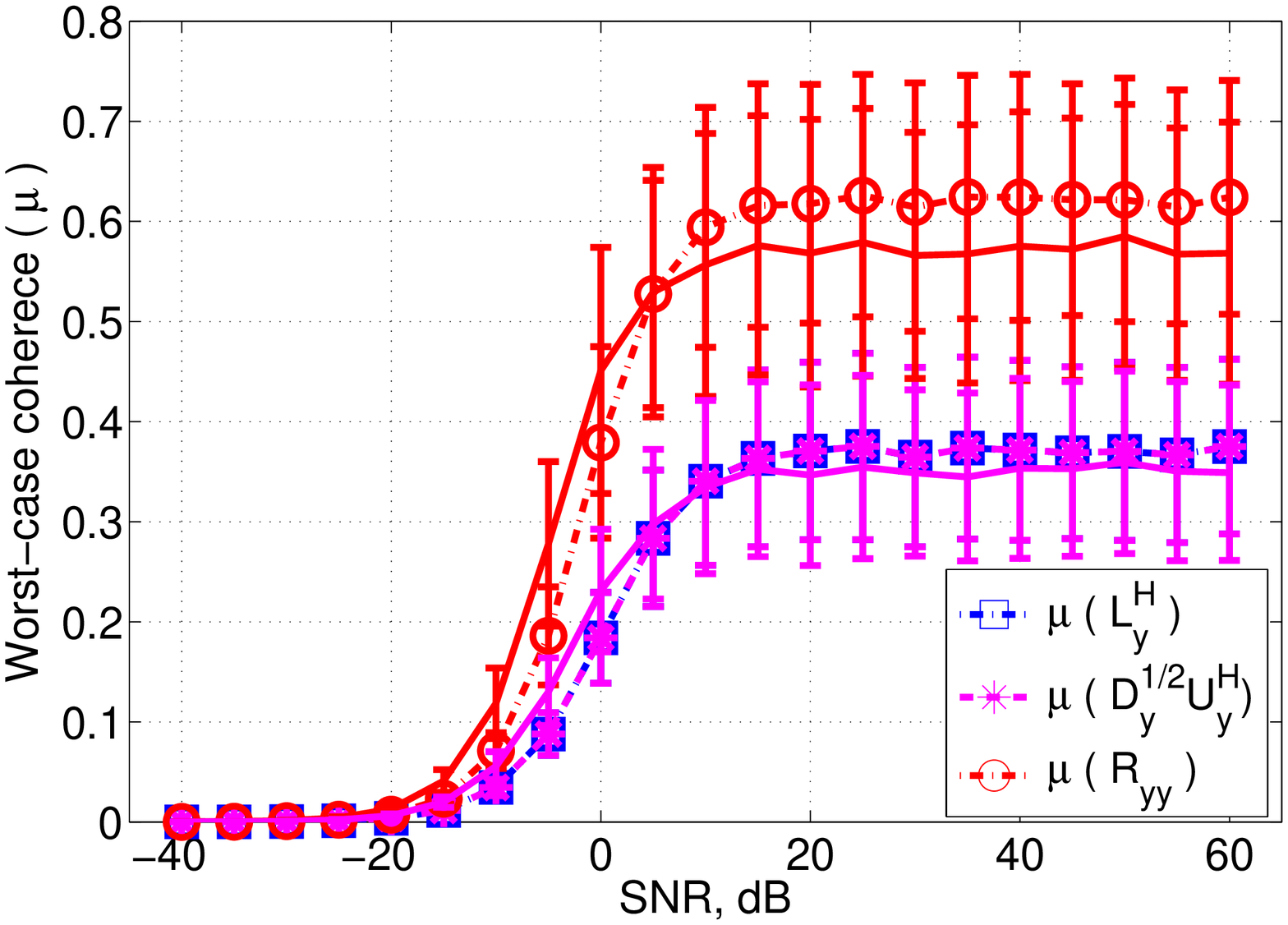}}

\textcolor{black}{\caption{{\footnotesize{}Worst-case coherence for the sparsifying dictionaries
$\boldsymbol{L}_{y}^{H}$, $\boldsymbol{D}_{y}^{\frac{1}{2}}\boldsymbol{U}_{y}^{H}$
and $\boldsymbol{R}_{yy}$ versus input SNR for UPDP with $n_{i}=2,\,n_{o}=2$,
$v=8$ and $N_{f}=80$. Solid lines represent the coherence of }\textcolor{black}{\footnotesize{}the
corresponding circulant approximation for $\boldsymbol{D}_{y}^{\frac{1}{2}}\boldsymbol{U}_{y}^{H}$
(i.e., $\boldsymbol{\Sigma^{H}}$) and }{\footnotesize{}$\boldsymbol{R}_{yy}$}\textcolor{black}{\footnotesize{}
(i.e., $\boldsymbol{\overline{R}}_{yy}=$$\boldsymbol{\Sigma}\boldsymbol{\Sigma^{H}}$).}\textcolor{black}{\label{fig:cohereneR_yy}}}
}
\end{figure}
\textcolor{black}{}
\begin{figure}[t]
\textcolor{black}{\includegraphics[scale=0.44]{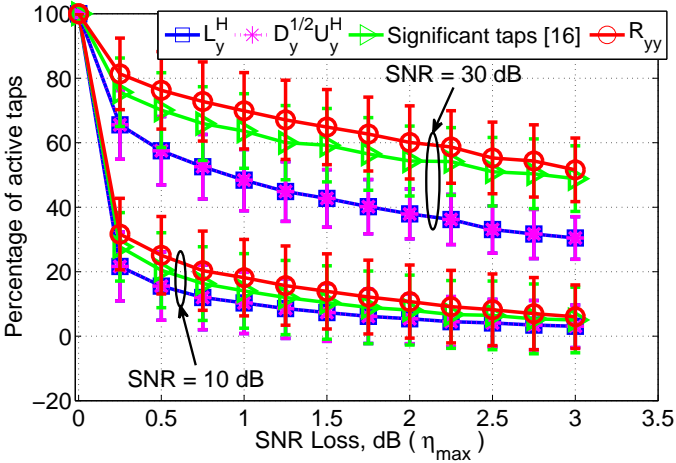}}

\textcolor{black}{\caption{{\footnotesize{}Percentage of active taps versus the performance loss
($\eta_{max}$) for the sparse MIMO-LEs with $\mbox{SNR (dB) \ensuremath{=10,\,30}}$,$\,n_{o}=2,\,n_{i}=2$,
$v=8$ and $N_{f}=80$.}\label{fig:activeTaps_versus_eta_max}}
}

\textcolor{black}{\vspace{-1.0em}}
\end{figure}
\textcolor{black}{}
\begin{figure}[t]
\textcolor{black}{\includegraphics[scale=0.43]{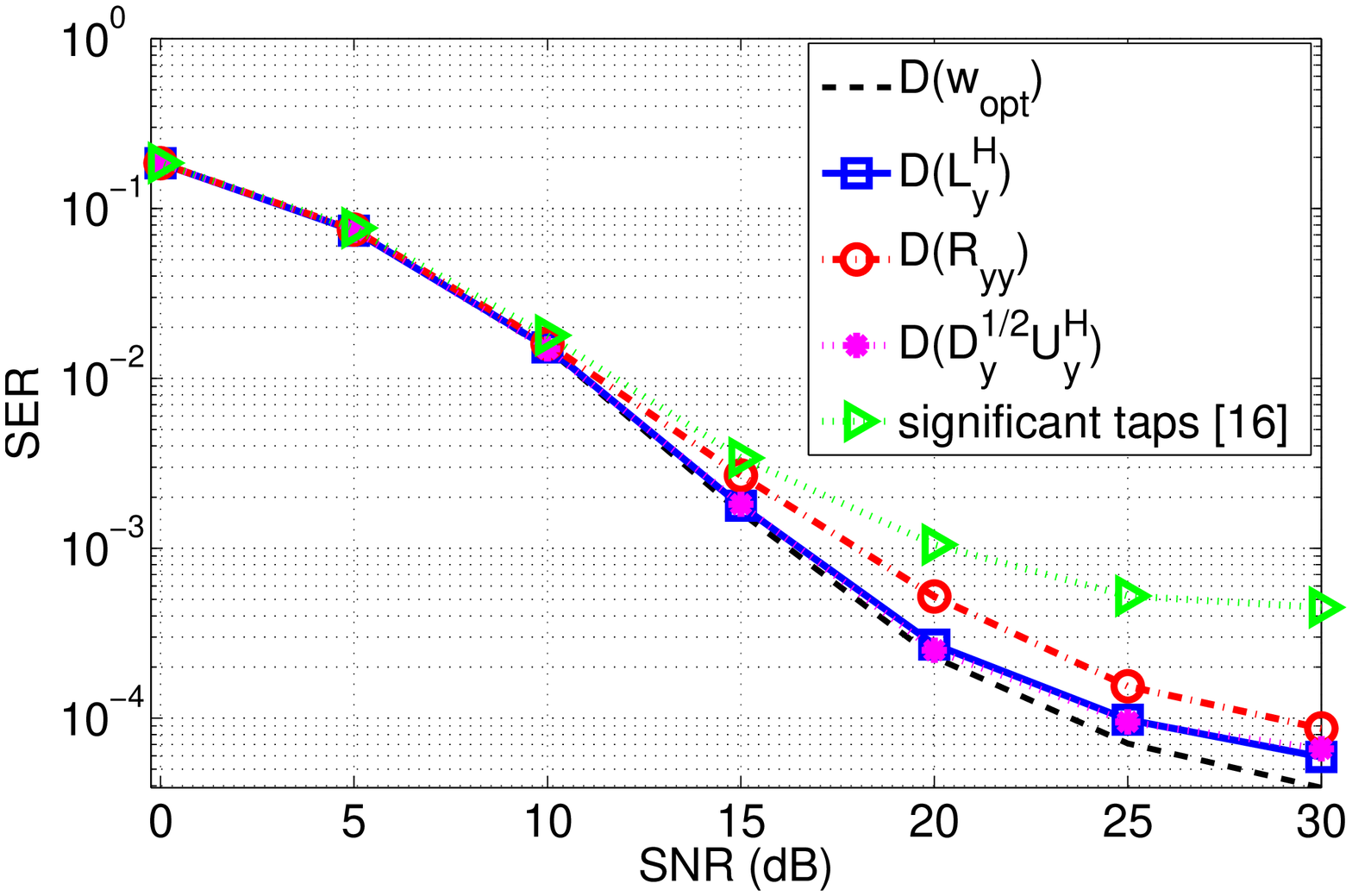}}

\textcolor{black}{\caption{{\footnotesize{}SER comparison between the non-sparse MMSE MIMO-LEs,
the proposed sparse MIMO-LEs $\boldsymbol{D}(\boldsymbol{D}_{y}^{\frac{1}{2}}\boldsymbol{U}_{y}^{H})$,
$\boldsymbol{D}(\boldsymbol{L}_{y}^{H})$, $\boldsymbol{D}(\boldsymbol{R}_{yy})$
and the ``significant taps'' based LE (proposed in \cite{sigTaps})
with sparsity level = 35\%, $n_{i}=2$, $n_{o}=2$, $v=5$ , $N_{f}=40$
and 16-QAM modulation.}\label{fig:BER_versus_SNR-1} }
}

\textcolor{black}{\vspace{-1.0em}}
\end{figure}
\textcolor{black}{{} }
\begin{figure}[t]
\textcolor{black}{\includegraphics[scale=0.43]{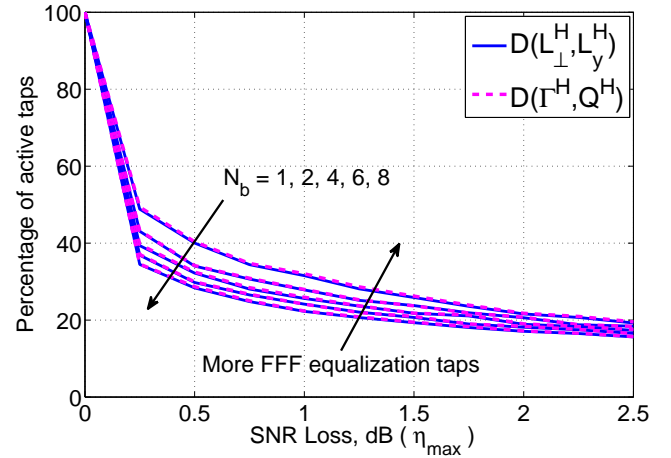}}

\textcolor{black}{\caption{\textcolor{black}{\footnotesize{}Percentage of active FFF taps versus
the performance loss ($\eta_{max}$) for sparse DFE designs with SNR
= 20 dB, $v=8$}\textcolor{black}{{} and $N_{f}=80$.\label{fig:siso-activeTaps-v-8nf-80}}}
}
\end{figure}
\textcolor{black}{}
\begin{figure}[t]
\textcolor{black}{\includegraphics[scale=0.43]{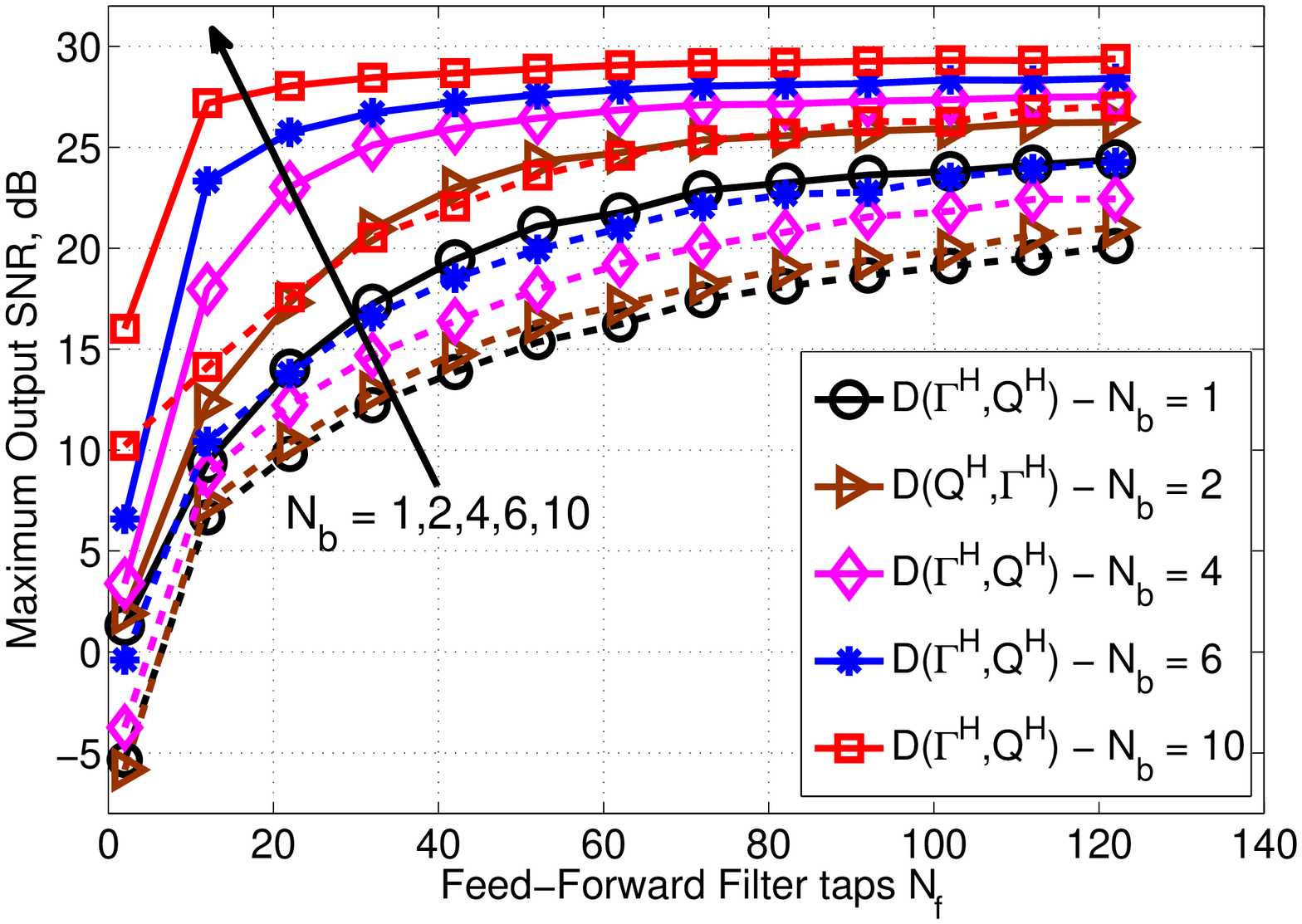}}

\textcolor{black}{\caption{\textcolor{black}{\footnotesize{}Maximum output SNR versus FFF taps
for UPDP CIR with }{\footnotesize{}$v=8$, $N_{f}=80$ and SNR = $30$
dB}\textcolor{black}{\footnotesize{}. Solid lines represent the }{\footnotesize{}$\boldsymbol{D}(\boldsymbol{\varGamma}^{H},\boldsymbol{Q}^{H})$
}\textcolor{black}{\footnotesize{}approach, while the dashed lines
represent the }{\footnotesize{}``significant taps'' approach proposed
in} {\footnotesize{}\cite{sigTaps}}.\textcolor{black}{{} \label{fig:unitTap-1}}}
}
\end{figure}
\textcolor{black}{}
\begin{figure}[t]
\textcolor{black}{\includegraphics[scale=0.43]{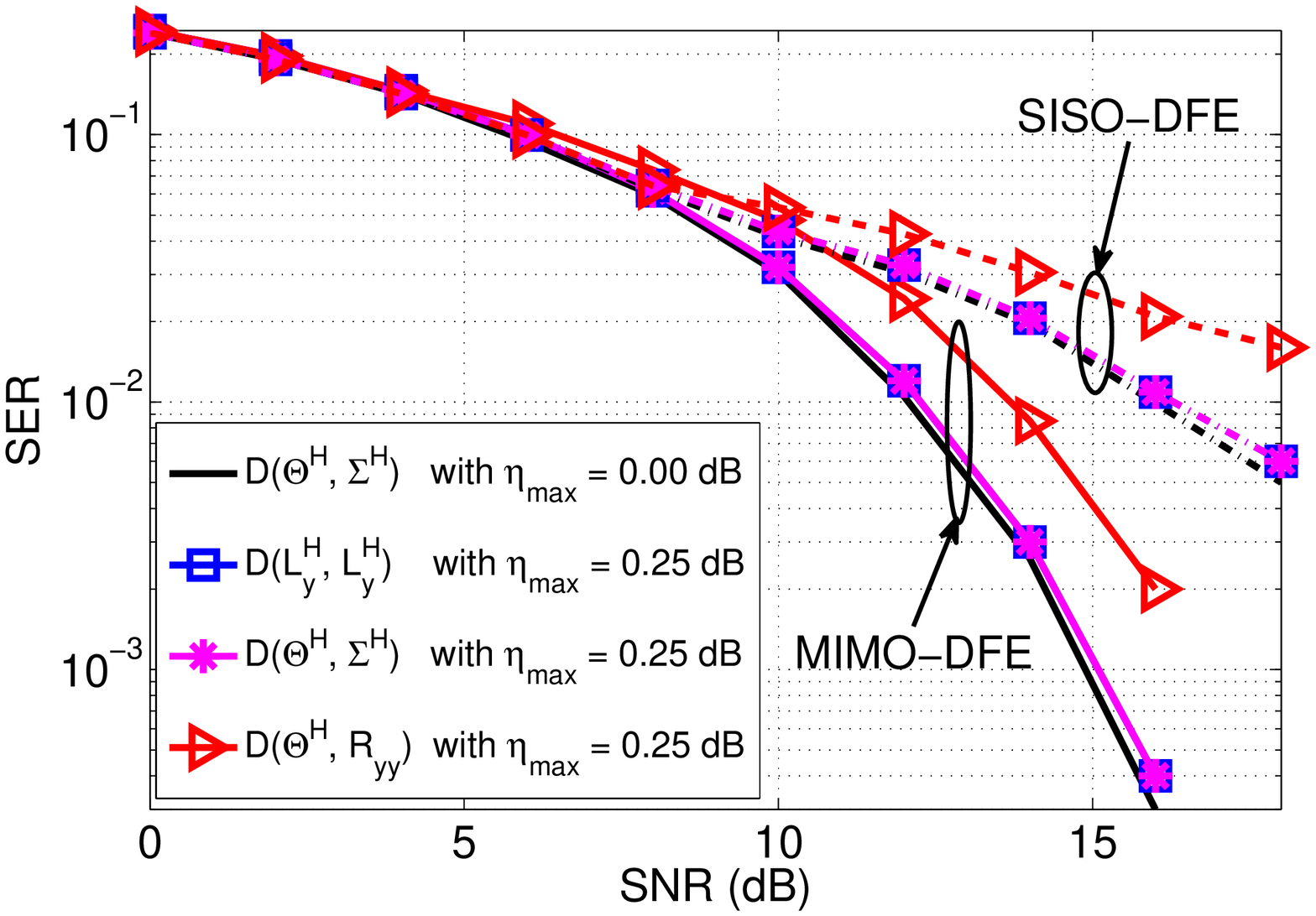}}

\textcolor{black}{\caption{{\footnotesize{}SER comparison between the MMSE non-sparse DFEs, i.e.,
$\boldsymbol{D}(\boldsymbol{\varTheta}^{H},\boldsymbol{\Sigma}^{H})$
with $\eta_{max}=0$, the proposed sparse DFEs $\boldsymbol{D}(\boldsymbol{L}_{y}^{H},\boldsymbol{L}_{y}^{H})$,
$\boldsymbol{D}(\boldsymbol{\varTheta}^{H},\boldsymbol{\Sigma}^{H})$
and the $\boldsymbol{D}(\boldsymbol{\varTheta}^{H},\boldsymbol{R}_{yy})$
for SISO and MIMO systems with $n_{i}=2,\,n_{o}=2$, $v=8$, $N_{b}=8$,
$N_{f}=80$, $\eta_{max}=0.25$ dB and 16-QAM modulation.}\label{fig:BER_versus_SNR}}
}
\end{figure}

\textcolor{black}{To investigate the coherence of the sparsifying
dictionaries used in our analysis, we plot the worst-case coherence
versus the input SNR in Figure \ref{fig:coherene } for sparsifying
dictionaries $\widetilde{\boldsymbol{L}}_{\perp}$ and $\boldsymbol{D}_{\perp}^{^{\frac{1}{2}}}\widetilde{\boldsymbol{U}}_{\perp}^{H}$
(which is formed by all columns of $\boldsymbol{D}_{\perp}^{\frac{1}{2}}\boldsymbol{U}_{\perp}^{H}$
except the $\left(\Delta+1\right)^{th}$ column) generated from $\boldsymbol{R}_{\perp}$.
Note that a smaller value of $\mu(\boldsymbol{\Phi})$ indicates that
a sparser approximation is more likely. Both sparsifying dictionaries
have the same $\mu\left(\boldsymbol{\varPhi}\right)$, which is strictly
less than $1$. Similarly, in Figure \ref{fig:cohereneR_yy}, we plot
the worst-case coherence of the proposed sparsifying dictionaries
used to design sparse MIMO-LEs and MIMO-DFEs. At high SNR levels,
the noise effects are negligible and, hence, the sparsifying dictionaries
(e.g., $\boldsymbol{R}_{yy}\approx\boldsymbol{H}\boldsymbol{H}^{H}$)
do not depend on the SNR. As a result, the coherence converges to
a constant. On the other hand, at low SNR, the noise effects dominate
the channel effects. Hence, the channel can be approximated as a memoryless
(i.e., 1 tap) channel. Then, the dictionaries (e.g., $\boldsymbol{R}_{yy}\approx\frac{1}{SNR}\boldsymbol{I}$)
can be approximated as a multiple of the identity matrix, i.e., $\mu\left(\boldsymbol{\varPhi}\right)\rightarrow0$.}

\textcolor{black}{Next, we compare different sparse FIR LE and DFE
designs based on different sparsifying dictionaries to study the effect
of $\mu\left(\boldsymbol{\varPhi}\right)$ on their performance. The
OMP algorithm is used to compute the sparse approximations. The OMP
stopping criterion is set to be a predefined sparsity level (number
of nonzero entries) or a function of the PRE such that: Performance
Loss ($\eta$)$=10\,\mbox{Log}_{10}\left(\frac{SNR(\widehat{\boldsymbol{z}}_{s})}{SNR(\boldsymbol{z}_{opt})}\right)\leq10\,\mbox{Log}_{10}\left(1+\frac{\epsilon}{\xi_{m}}\right)\triangleq\eta_{max}$.
Here, $\epsilon$ is computed based on an acceptable $\eta_{max}$
and, then, the coefficients of $\widehat{\boldsymbol{z}}_{s}$ are
computed through (\ref{eq:propFW}). The percentage of the active
taps is calculated as the ratio between the number of nonzero taps
to the total number of filter taps. For the MMSE equalizer, where
none of the coefficients is typically zero, the number of active filter
taps is equal to the filter span. The decision delay for LEs is set
to be $\Delta\approx\frac{N_{f}+v}{2}$ \cite{jcioffi}, while for
DFEs we set $\Delta\approx N_{f}-1$, which is optimum when $N_{b}=v$
\cite{MMSE-DFE}.}

\textcolor{black}{Figure \ref{fig:activeTaps_versus_eta_max} plots
the percentage of the active taps versus the performance loss $\eta_{max}$
for the proposed sparse FIR MIMO-LEs and the proposed approach in
\cite{sigTaps}, which we refer to it as the ``significant taps''
approach. In that approach, all of the FIR filter taps are computed
and only the $\nu$-significant ones are retained. We observe that
a lower active taps percentage is obtained when the coherence of the
sparsifying dictionary is small. For instance, allowing for $0.25$
dB SNR loss results in a significant reduction in the number of active
LE taps. Approximately two-thirds (respectively, two-fifths) of the
taps are eliminated when using $\boldsymbol{D}_{y}^{\frac{1}{2}}\boldsymbol{U}_{y}^{H}$
and $\boldsymbol{L}_{y}^{H}$ at SNR equal to 10$\,$(respectively,
30). The sparse MIMO-LE designed based on $\boldsymbol{R}_{yy}$ needs
more active taps to maintain the same SNR loss as that of the other
sparse MIMO-LEs due to its higher coherence. This suggests that the
smaller the worst-case coherence of the dictionary in our setup, the
sparser is the equalizer. Moreover, a lower sparsity level (active
taps percentage) is achieved at higher SNR levels, which is consistent
with the previous findings (e.g., in \cite{tapPositions07}). Furthermore,
reducing the number of active taps decreases the filter equalization
design complexity and, consequently, power consumption since a smaller
number of complex multiply-and-add operations are required. }

\textcolor{black}{In Figure \ref{fig:BER_versus_SNR-1}, we compare
the symbol error rate (SER) performance of our proposed sparse FIR
MIMO-LEs with the ``significant taps'' approached proposed in \cite{strongestTap}.
Assuming a $25\%$ sparsity level, both the $\boldsymbol{D}(\boldsymbol{D}_{y}^{\frac{1}{2}}\boldsymbol{U}_{y}^{H})$
and $\boldsymbol{D}(\boldsymbol{L}_{y}^{H})$ sparse LEs achieve the
lowest SER followed by $\boldsymbol{D}(\boldsymbol{R}_{yy})$, while
the ``significant taps'' performs the worst. In addition to this
performance gain, the complexity of the proposed sparse LEs is less
than that of the ``significant-taps'' LE since only an inversion
of an $N_{s}\times N_{s}$ matrix is required (not $N_{f}\times N_{f}$
as in the ``significant-taps'' approach) where $N_{s}$ is the number
of nonzero taps. Although the $\boldsymbol{D}(\boldsymbol{D}_{y}^{\frac{1}{2}}\boldsymbol{U}_{y}^{H})$
and $\boldsymbol{D}(\boldsymbol{L}_{y}^{H})$ LEs achieve almost the
same SER, the former has a lower decomposition complexity since its
computation can be done efficiently using only the FFT and its inverse. }

\textcolor{black}{The effect of our sparse FFF and FBF FIR filter
designs for SISO DFEs on the performance is shown in Figure \ref{fig:siso-activeTaps-v-8nf-80}.
We plot the active (non-zero) FFF taps percentage of the total FFF
span $N_{f}$ versus the maximum loss in the output SNR. Allowing
a higher loss in the output SNR yields a bigger reduction in the number
of active FFF taps. Moreover, the active FFF taps percentage increases
as $N_{b}$ decreases because the equalizer needs more taps to equalize
the CIR. We also observe that allowing a maximum of only }\textbf{\textcolor{black}{$0.25$}}\textcolor{black}{{}
dB in SNR loss with $N_{b}=4$ results in a substantial $60\%$ reduction
in the number of FFF active taps (the equalizer can equalize the channel
using only 32 out of 80 taps). }

\textcolor{black}{In Figure \ref{fig:unitTap-1}, we compare our proposed
sparse FBF design with that in \cite{sigTaps}, i.e., the ``significant
taps'' approach, in terms of output SNR where we plot the output
SNR versus FFF taps $N_{f}$ for the UPDP channel. We vary $N_{b}$,
the number of FBF taps, from $1$ (lower curve) to $10$ (upper curve).
The output  SNR increases as $N_{b}$ increases for all FBF designs,
as expected, and our sparse FBF outperforms, for all scenarios, the
proposed approach in \cite{sigTaps}. Notice that as $N_{b}$ increases,
the sparse FBF becomes more efficient in removing ISI from previously-detected
symbols resulting in a higher SNR.}

\textcolor{black}{In Figure \ref{fig:BER_versus_SNR}, we study the
SER performance of our proposed sparse SISO-DFEs and MIMO-DFEs versus
the input SNR based on different design criteria and using different
sparsifying dictionaries. Assuming a maximum SNR loss of $0.25$ dB,
both $\boldsymbol{D}(\boldsymbol{L}_{y}^{H},\boldsymbol{L}_{y}^{H})$
and $\boldsymbol{D}(\boldsymbol{\varTheta}^{H},\boldsymbol{\Sigma}^{H})$
sparse SISO/MIMO DFEs designs achieve the lowest SER, followed by
$\boldsymbol{D}(\boldsymbol{\varTheta}^{H},\boldsymbol{R}_{yy})$.
Note that $\eta_{max}=0$ corresponds to the optimum non-sparse MMSE
design where all the equalizer taps are active. Additionally, at high
SNR, diversity gains of the MIMO-DFEs over the SISO-DFEs are noticed
resulting in a better performance. }

\section{\textcolor{black}{Conclusions\label{sec:Conclusion-and-Future}}}

\textcolor{black}{In this paper, we proposed a general framework for
designing sparse FIR MIMO LEs and DFEs based on a sparse approximation
formulation using different dictionaries. Based on the asymptotic
equivalence of Toeplitz and circulant matrices, we also proposed reduced-complexity
designs, for all proposed FIR filters, where matrix factorizations
can be carried out efficiently using the FFT and inverse FFT with
negligible performance loss as the number of filter taps increases.
In addition, we analyzed the coherence of the proposed dictionaries
involved in our design and showed that the dictionary with the smallest
coherence gives the sparsest filter design. Finally, the significance
of our approach was shown analytically and quantified through simulations. }

\bibliographystyle{IEEEtran}
\bibliography{sparseFIR_MIMO_Ref}

\end{document}